\documentclass[twocolumn]{aastex631}

\usepackage{amsmath}
\usepackage{mathtools}
\usepackage{amssymb}
\usepackage{amsfonts}
\usepackage{bm}
\usepackage{graphicx}
\usepackage{xspace}

\newcommand{\IRSSquare}{IRS$^2$\xspace}

\newcommand{\ie}{{\it i.e.}\xspace}

\newcommand{\eg}{{\it e.g.}\xspace}

\newcommand{\etal}{{\it et al.}\xspace}

\xspace

\newcommand{\HST}{{\em HST}\xspace}
\newcommand{\JWST}{{\em JWST}\xspace}
\newcommand{\ACE}{{\em ACE}\xspace}
\newcommand{\GOES}{{\em GOES}\xspace}

\newcommand{\RomanST}{{\em Roman}\xspace}

\usepackage{xcolor}

\newif\ifhighlight
\highlighttrue  

\newcommand{\myedit}[1]{%
  \ifhighlight
    {\color{red}\bfseries #1}
  \else
    #1
  \fi
}

\newcommand{\xmyedit}{\highlightfalse}

\xmyedit

\begin{document}

\title{JWST NIRSpec's Cosmic Ray Experience at L2\footnote{\raggedright Published in PASP, 137, 095003 (2025); DOI: 10.1088/1538-3873/adfb2f}}

\correspondingauthor{Bernard J. Rauscher}
\email{Bernard.J.Rauscher@nasa.gov}

\author[0000-0003-2662-6821]{Bernard J. Rauscher}
\affiliation{NASA Goddard Space Flight Center, Observational Cosmology Laboratory, Greenbelt, MD 20771, USA}

\author{D.J. Fixsen}
\affiliation{NASA Goddard Space Flight Center, Observational Cosmology Laboratory, Greenbelt, MD 20771, USA}
\affiliation{Center for Research and Exploration in Space Science and Technology, NASA/GSFC, Greenbelt, MD 20771, USA}
\email{Dale.J.Fixsen@nasa.gov}

\begin{abstract}
We characterize cosmic ray interactions in blanked-off \JWST NIRSpec ``dark'' exposures. In its Sun/Earth-Moon L2 halo orbit, \JWST encounters energetic ions that penetrate NIRSpec's radiation shielding. The shielded cosmic ray hit rate decreased from approximately $4.3$ to $2.3~\mathrm{ions~cm^{-2}}~s^{-1}$ during the first three years of operation. A typical hit affects about 7.1~pixels necessitating mitigation during calibration and deposits around $6~\mathrm{keV}$ in the $\lambda_\mathrm{co} = 5.4~\mu$m HgCdTe detector material (equivalent to $\sim5200$ charges). The corresponding linear energy transfer is about $0.86~\mathrm{keV~\mu m^{-1}}$. As we are currently near solar maximum, galactic cosmic ray flux is expected to increase as solar activity declines, leading to an anticipated rise in the NIRSpec rate from $2.3$ to $4.3~\mathrm{ions~cm^{-2}}~s^{-1}$ by early 2027 and potentially reaching $\sim6~\mathrm{ions~cm^{-2}}~s^{-1}$ in the early 2030s. We investigate rare, large ``snowball'' hits and, less frequently, events with secondary showers that pose significant calibration challenges. We explore their possible origins as heavy ions, secondary particles from shielding, or inelastic scattering in the HgCdTe detector material. We discuss the implications of these findings for future missions including the Nancy Grace Roman Space Telescope.
\end{abstract}

\section{Introduction}\label{sec:intro}

The James Webb Space Telescope \citep[\JWST;][]{Gardner2023} was launched on December 25, 2021, and began scientific observations the following June. Since then, \JWST's science instruments have regularly collected blanked-off ``dark'' 
exposures for calibration purposes. These darks contain a wealth of information about the Near Infrared Spectrograph (NIRSpec) and its detectors' response to space radiation. While NIRSpec's radiation shielding blocks most low energy particles, high energy galactic cosmic rays (GCR) pass through. They leave ionization tracks behind that manifest mostly as small clusters of disturbed pixels or streaks. These cosmic ray ``hits'' are an important consideration in observation planning. Detecting and correcting for them is an important step in the calibration pipeline.

The purpose of this article is to describe the appearance of NIRSpec cosmic rays from the perspective of astronomers. Our focus is on parameters and disturbances that affect observation planning and scientific data integrity. However, space radiation poses other concerns from a hardware safety and operation perspective. These effects include long-term performance degradation, single event upsets, and spacecraft charging. For more information on these topics, the interested reader is referred to \citet{Barth:2000vc} and \citet{Evans2003}.

\myedit{One example of performance degradation is the steady increase of inoperable ``hot'' pixels in the NIRSpec detectors. Since June, 2022, the rate of increase has been about 100 new hot pixels per month for each of the two detectors, or about 0.03\% of all pixels per year.}\protect\footnote{\url{https://jwst-docs.stsci.edu/jwst-near-infrared-spectrograph/nirspec-instrumentation/nirspec-detectors/nirspec-detector-performance\#NIRSpecDetectorPerformance-Badpixels}}

The remainder of this paper is organized as follows. Section~\ref{sec:jwst} provides more information on \JWST. We then describe the data used in this study, including pipelined calibration and the cosmic ray finder algorithm, in Section~\ref{sec:data}. We present our results in Section~\ref{sec:results}. Section~\ref{sec:other-missions} compares the cosmic ray rate observed by NIRSpec with that of other concurrent missions. Looking ahead, Section~\ref{sec:roman} discusses our expectations for the Nancy Grace Roman Space Telescope (\RomanST). Finally, Section~\ref{sec:summary} is the summary.

\section{JWST}\label{sec:jwst}

\JWST is today's premier space observatory for $0.6-28~\mu$m astrophysics. It features an approximately 6.5~m diameter cryogenic space telescope and four scientific instruments: (1) NIRSpec, (2) the Near Infrared Camera (NIRCam), (3) the Near Infrared Imager and Slitless Spectrograph (NIRISS), and (4) the Mid-Infrared Instrument (MIRI). The instruments' detectors are exquisitely sensitive to cosmic rays in addition to light. The observatory is situated in a halo orbit around the Sun/Earth-Moon system's second Lagrange point (L2).

Since June, 2022, NIRSpec has collected blanked-off dark exposures at regular intervals. These darks reveal the cosmic ray environment at NIRSpec's detectors. Although the natural L2 environment provides the cosmic rays, NIRSpec's detectors sit behind radiation shielding. The shielding blocks most low energy particles ($\mathrm{E \lesssim 100~MeV}$). However, it is possible for low energy secondary radiation to be produced within the shielding itself. High energy GCRs easily penetrate the shielding.

We selected NIRSpec for this study because we are part of the team that built it. This provides better insight into the radiation shielding than we have for other instruments. Alone among \JWST instruments, NIRSpec provides Improved Reference Sampling and Subtraction \citep[\IRSSquare;][]{Rauscher_2017}. \IRSSquare uses a specialized detector clocking pattern and reference correction software to significantly suppress correlated read noise compared to NIRCam and NIRISS. Using \IRSSquare, cosmic ray detection is less affected by detector noise than it would be for other instruments.

This section builds on two early studies of the L2 space environment that were done for \JWST. At the time, the concept for a large, IR-optimized space telescope, was known as the Next Generation Space Telescope (NGST). Our aim here is to provide enough information about the environment for readers to understand later sections of this paper in context. For a more complete explanation of the L2 environment, we refer the interested reader to the original studies \citep{Barth:2000vc,Evans2003}. 

\subsection{L2 Halo Orbit and Radiation Environment}\label{sec:L2}

\begin{figure}
\center
\includegraphics[width=\columnwidth]{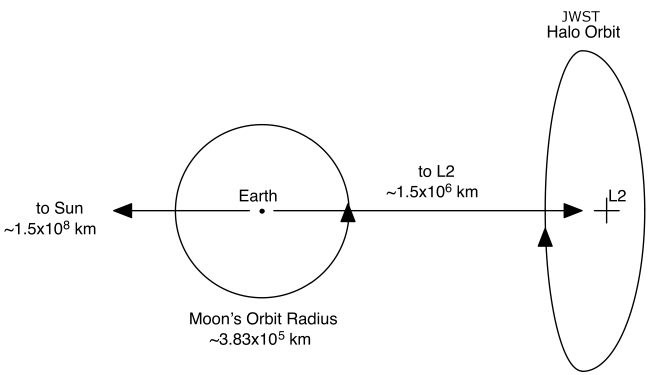}
\caption{\JWST is in a ``halo'' orbit about the Sun/Earth-Moon L2 Lagrange point. L2 is about $1.5\times10^6~\mathrm{km}$ from the earth, outside the Moon's orbit, and on the opposite side of the Earth from the Sun. Credit: Based on a figure from \citep{Evans2003}}\label{fig:l2_orbit}
\end{figure}

Figure~\ref{fig:l2_orbit} shows \JWST's orbit in relation to the Sun and the Earth-Moon gravitational system. L2 is located approximately $1.5 \times 10^6$~km from 
Earth on the anti-Sun side, along the line joining the Sun and the Earth-Moon center of mass. This orbit was chosen because it allows a single sunshade to 
shield the observatory from the Sun, Earth, and Moon, thereby facilitating passive cooling. Moreover, \JWST's orbit enables observing nearly half of the sky at any 
given time while maintaining acceptable radio contact with Earth.

Typically, a spacecraft in a heliocentric orbit with a radius larger than Earth's 
moves at a slower angular velocity, resulting in a longer orbital period and 
causing it to lag behind Earth. However, at the L2 point, the combined 
gravitational pull of the Earth and Moon provides the necessary additional 
acceleration for \JWST to maintain orbital velocity commensurate with Earth's solar orbit, thus keeping pace. The L2 halo orbit is dynamically unstable; small gravitational perturbations from other bodies tend to nudge \JWST away from L2 into an independent solar orbit. Therefore, periodic station-keeping maneuvers are required to maintain \JWST's position within the halo orbit.

\begin{figure*}
\center
\includegraphics[width=\textwidth]{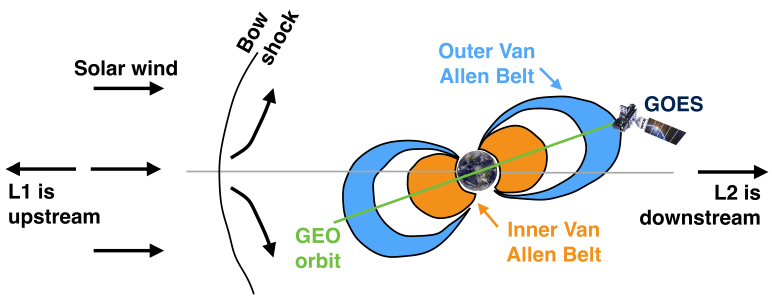}
\caption{Noon-midnight cross section of the near-Earth environment. Magnetosphere plasma extends out to about the bow shock looking toward the sun and trails further downstream on the anti-Sun side. \JWST's L2 halo orbit is about $235~\mathrm{R}_\earth$ to the right of the earth. 
This is deep space, where the earth's magnetosphere provides essentially no 
shielding. The Advanced Composition Explorer (\ACE) is located at L1. This is also 
deep space. We would expect \ACE to experience a similar GCR environment to \JWST.
The \GOES satellites are in geostationary orbits. There, the Earth's magnetic 
shielding provides some protection. We therefore expect \JWST and \ACE to see 
similar GCRs. One might reasonably expect \GOES to see somewhat fewer GCRs on 
account of the weak, but still present protection from the earth's magnetosphere.}\label{fig:magnetosphere}
\end{figure*}

The Earth’s liquid outer core generates an approximately dipole magnetic field that dominates the magnetosphere within a few Earth radii. Figure~\ref{fig:magnetosphere} shows the magnetic field and plasma environment near the earth. The magnetic field deflects charged particles and thus ``protects'' the Earth from some radiation. However, \JWST operates in deep space, beyond this protection.

Figure~\ref{fig:magnetosphere} shows the locations of \JWST and two other missions
that are discussed in Section~\ref{sec:other-missions}. \JWST is located far to 
the 
right of this figure, where the magnetosphere provides essentially no shielding.
It is in deep space. \ACE is located at L1, 
well to the left of this figure. It too is in deep space. The Geostationary 
Operational Environmental Satellites (\GOES) are in geostationary orbits (GEO). 
The thick, yellow line in Figure~\ref{fig:magnetosphere} shows the locations of GEO 
orbits. The \GOES satellites are located outside the Inner Van Allen Belt, but 
near the outer edge of the Outer Van Allen Belt.

The Outer Van Allen Belt is dominated by electrons that are blocked by NIRSpec's radiation shielding. Although the Earth's magnetic field is weak at GEO ($\sim100$~nT),\footnote{See the GOES Magnetometer webpage at \url{https://www.swpc.noaa.gov/products/goes-magnetometer}. $Bt$ is the magnetic field strength.} it is nevertheless significantly stronger than the $\lesssim10$~nT Interplanetary Magnetic Field (IMF) that dominates at L2.\footnote{See the ACE real time solar wind webpage at \url{https://www.swpc.noaa.gov/products/ace-real-time-solar-wind} where $Bt$ is the magnetic field strength.} Bearing these considerations in mind, we expect \ACE's GCR environment to be similar to \JWST's. We expect \GOES to be exposed to a somewhat weaker GCR environment than \JWST on account of the weak but still present shielding provided by the magnetosphere. Figure~\ref{fig:magnetosphere} shows several other magnetosphere features that do not concern us here. For more information, the interested reader is referred to \citet{Evans2003}.

The Sun generates a solar wind of charged particles that carry a magnetic
field out from the Sun.  As the magnetic field spreads, it weakens until it 
balances the magnetic pressure of interstellar space at about 150 AU.
The space within, known as the heliosphere, is dominated by the solar wind
and the magnetic fields that it carries.

Charged  particles from the Sun are mostly low energy protons (up to $\sim$50~Mev)
which are shielded from the detectors by material around the detectors.
GCRs from beyond the solar system are dominated ($\sim$90\%)
by few GeV protons. The remaining cosmic particles are 
alpha particles ($\sim$ 9\%) with a few heavier elements and exotic particles.
The solar magnetic field affords some protection from these particles 
but the solar magnetic field and thus the shielding fluctuates with the 
solar cycle. Cosmic radiation levels are high when solar activity 
and sunspots are low and vice versa.

\subsection{NIRSpec}\label{sec:nirspec}

Our discussion of NIRSpec is limited to components that affect its
sensitivity to cosmic rays. These are primarily the H2RG near infrared array detectors and the radiation shielding. For a full description of the instrument, the interested reader is referred to \citet{Jakobsen_2022}. \citet{Birkmann2022} and references therein provide more information about NIRSpec's detectors.

\begin{figure}
\center
\includegraphics[width=.66\columnwidth]{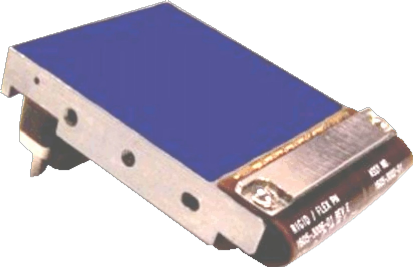}
\caption{NIRSpec has two Teledyne H2RG HgCdTe near infrared detector arrays. Light enters from the top. Cosmic rays come in from all sides. Those that
come through the bottom pass through about $\approx$7~mm of molybdenum, a thin balanced composite structure (BCS), thin layers of epoxy, a thin silicon readout
integrated circuit (ROIC), and thin indium bumps before reaching the HgCdTe 
detector layer. The molybdenum (Mo) package is included in the shielding on the
back side of the detector. The BCS is a proprietary component that Teledyne uses to minimize thermally induced strain. Everything except the molybdenum base is
ignored in the shielding calculation.}\label{fig:h2rg}
\end{figure}

NIRSpec has two Teledyne H2RG HgCdTe near infrared detector arrays 
(Figure~\ref{fig:h2rg}). When a photon hits the HgCdTe detector material, it 
typically creates an electron-hole pair. The holes are collected in pixels which 
charge the photodiodes' capacitances. NIRSpec's near-IR detector arrays thus convert photons into 
voltages in pixels. Then, the SIDECAR application specific integrated circuit 
(ASIC) readout electronics amplify and convert these voltages into digital numbers 
(DN) in four separate output channels for each detector. 

It is standard practice in the astronomical community to parameterize near-IR array detectors in units of electrons. Although NIRSpec's p-on-n photododiodes physically integrate holes, we nevertheless follow convention from this point forward and speak of electrons except when the difference is physically important to the discussion at hand. The conversion gains 
between electrons and DN, $g_c~ (e^-/\mathrm{DN})$, were measured at NASA Goddard 
Space Flight Center as part of characterizing the NIRSpec Detector Subsystem (Table~\ref{tab:pp-gain}). 

The pair creation energy is set by the HgCdTe bandgap energy, $E_g$. This is related to the cutoff wavelength by,
\begin{equation}
E_g=\frac{h c}{\lambda_\mathrm{co}}.\label{eq:pp-energy}
\end{equation}
Teledyne measures $\lambda_\mathrm{co}$ when building detectors. \citet{Alig1977} provides an empirical relation for converting the bandgap energy of a semiconductor to pair-creation energy, $\epsilon$. It is,
\begin{equation}
\epsilon = 2.73 E_g + 0.55~\mathrm{eV}.\label{eq:alig}
\end{equation}
\citet{Fox2009} validated this expression for HgCdTe using data from eight non-flight Hubble Space Telescope (\HST) Wide Field Camera 3 (WFC3) Teledyne H1R arrays. Substituting Equation~\ref{eq:pp-energy} into Equation~\ref{eq:alig} and multiplying by $g_c$ yields the relation,
\begin{equation}
g_\mathrm{pp} = \left(2.73\frac{h c}{\lambda_\textrm{co} e}+0.55\right)g_c\hspace{18pt}\mathrm{eV/DN}.\label{eq:pp_gain}
\end{equation}
Table~\ref{tab:pp-gain} evaluates Equation~\ref{eq:pp_gain} for the two NIRSpec detectors. For more information about NIRSpec's H2RG detectors, the reader is referred to \citet{Rauscher2014}. For more information about H2RG detectors in general, we refer the interested reader to \citet{Loose:2003vh,Loose2007}.

\begin{deluxetable}{cccccc}
    \tablecaption{Pair-production Gains}\label{tab:pp-gain}
    \tablehead{\colhead{}& \colhead{}& \colhead{$\lambda_\mathrm{co}$}& \colhead{Output}& \colhead{$g_c$}& \colhead{$g_\mathrm{pp}$}\\
    \colhead{Channel}& \colhead{SCA}& \colhead{($\mu$m)}& \colhead{(\#)}& \colhead{($e^-/\mathrm{DN}$)}& \colhead{(eV/DN)}}
    \startdata
        NRS1& 17163& 5.45& 1& 0.990& 1.159\\
           &      &     & 2& 0.989& 1.158\\
           &      &     & 3& 1.001& 1.172\\
           &      &     & 4& 1.005& 1.176\\
           &      &     & Mean& 0.996& 1.167\\[2ex]
        NRS2& 17280& 5.37& 1& 1.119& 1.320\\
           &      &     & 2& 1.123& 1.325\\
           &      &     & 3& 1.142& 1.347\\
           &      &     & 4& 1.165& 1.375\\
           &      &     & Mean& 1.137& 1.342
    \enddata
\end{deluxetable}

\subsection{NIRSpec Radiation Shielding}\label{sec:shielding}

Radiation shielding is provided by the molybdenum (Mo) focal plane assembly (FPA) structure, the silicon carbide (SiC) camera housing, various SiC mirrors, and the SiC optical bench. Roughly speaking, there are about 20~mm of SiC shielding  on the side of the detectors that light enters from. There are about 12~mm of Mo shielding on the back side. Shielding on the back side includes the Mo pedestal visible in Figure~\ref{fig:h2rg} and the Mo focal plane assembly itself.

The low energy electrons in the solar wind do not penetrate the science instruments, much less the engineered radiation shielding. The National Institute of Standards and Technology (NIST) Stopping Power and Range Tables for Protons 
(PSTAR)\footnote{\url{https://physics.nist.gov/PhysRefData/Star/Text/PSTAR.html}} program calculates stopping power and range tables for protons in various materials. Using PSTAR, we calculate that this shielding blocks essentially all 
solar protons during solar-calm periods. It also blocks GCRs having energy $\lesssim 100$~MeV.

\subsection{Pre-Launch Expectations}\label{sec:expectations}

Several studies were done before launch to understand the cosmic ray environment
at L2 and how NIRSpec's detectors might respond to it, including 
\citet{Evans2003} that describes the natural environment at L2, with a short
discussion of radiation effects. \citet{Barth:2000vc} focused specifically on the 
radiation environment and provided the pre-launch ``nominal'' cosmic ray rate 
estimate of $5~\mathrm{ions~cm^{-2}}~s^{-1}$. \citet{Pickel2002} used physical 
modeling to study how semiconductor detector arrays respond to ionizing 
radiation. Pickel's work was done before the flight detector designs were 
finalized. Although Pickel~\etal included physical models of charge spreading, 
it could not simulate flight-like \JWST detectors because the design parameters were not yet known.

As high energy ions traverse the HgCdTe detectors they leave trails of 
holes that are collected by the pixels. Here we deviate from the convention established in Section~\ref{sec:nirspec} because it is important to understand that holes are being collected. These holes integrate in the photodiodes and are then 
read out with other holes freed by photons from the telescope.
Minimum ionizing protons (MIP), around 2~GeV, deposit about 1~keV/$\mu$m as 
they travel through the detector.  Since the rise of energy deposit
is only logarithmic with the rise of the proton energy, this minimum
is a reasonable approximation for typical cosmic rays.  The detector
thickness is $\sim$5~$\mu$m leading to a prediction of $\sim$5~keV for protons
normal to the detectors.

\citet{Giardino2019} examined how mitigating cosmic rays in the pipeline would affect NIRSpec sensitivity. They used a library of simulated cosmic ray 
hits to study the effect of the planned pipeline cosmic ray mitigation on NIRSpec 
noise. Giardino~\etal injected theoretical cosmic rays into real NIRSpec ground 
test data. They predicted that overall effect of cosmic ray mitigation by the 
pipeline would increase NIRSpec's noise by about 7\%. \citet{Birkmann2022} studied NIRSpec's total noise during commissioning and found that the measured values were consistent with pre-flight predictions.

\section{Data}\label{sec:data}

\subsection{NIRSpec Darks}

The data come from three ``Full Frame Dark Monitor'' programs. The Program IDs are 1484, 4455, and 6633 in Cycles 1, 2, and 3 respectively. The files are available from the Mikulski Archive for Space Telescopes (MAST)\footnote{\url{https://archive.stsci.edu/}} at the Space Telescope Science Institute (STScI). We used all available \IRSSquare darks from these programs through 13 February 2025. The integrations use NIRSpec's NRSIRS2RAPID mode. This resets the detector pixel-by-pixel, and then acquires 200 non-destructive samples up-the-ramp at a constant $t_f=14.58889$ seconds per frame cadence. Nearly every pixel is disturbed by a cosmic ray during the roughly 49 minute integration time.

\subsection{Up-the-ramp Readout}

NIRSpec uses a sampling ``up-the-ramp'' readout scheme. Each $2048\times 2048$~pixel detector is read out in four $512\times 2048$~pixel stripes. Both detectors and all stripes are clocked synchronously. Within a stripe, there are 512 pixels in the fast-scan direction and 2048 lines in the slow-scan direction. The detectors are clocked at a constant $t_f$ frame cadence. Destructive reset frames and non-destructive read frames have identical pixel timing.

In up-the-ramp sampling, each integration begins with a reset frame. Then, the system acquires the desired number of non-destructive reads. When there is light on the detector, a plot of the signal in a pixel versus time increases approximately linearly (``up-the-ramp'') until saturation is approached. More information about NIRSpec's clocking patterns can be found in the online \JWST User Documentation.\footnote{\url{https://jwst-docs.stsci.edu/jwst-near-infrared-spectrograph/nirspec-instrumentation/nirspec-detectors/nirspec-detector-readout-modes-and-patterns}}

\subsection{JWST Pipeline Processing}

We used Stage 1 of the \JWST pipeline to process the uncalibrated data (filename suffix \texttt{\_uncal.fits}).\footnote{\url{https://jwst-docs.stsci.edu/jwst-science-calibration-pipeline/stages-of-jwst-data-processing/calwebb_detector1}} The pipeline is readily configurable giving users the option of choosing which steps to apply. We used the default settings in pipeline version 1.13.4 to apply \texttt{dq\_init\_step}, \texttt{saturation\_step}, \texttt{superbias\_step}, \texttt{refpix\_step}, and \texttt{linearity\_step} to all of the \texttt{\_uncal.fits} files.

The first two steps are housekeeping. Then, \texttt{superbias\_step} removes the approximately constant detector bias pattern. For NIRSpec, \texttt{refpix\_step} applies the \IRSSquare reference correction. Finally, \texttt{linearity\_step} corrects for classical, signal-dependent nonlinearity. The resulting \texttt{\_linearitystep.fits} files are the input to our own cosmic ray finder.

\subsection{Cosmic Ray Finder Algorithm}\label{sec:finder}

To have as much control as possible, we wrote our own cosmic ray finder (\ie we did not use the pipeline's \texttt{jump\_step}). First, data is processed to find the typical (RMS) noise and dark current in each pixel. A few ($<$3\%) pixels are excluded from the study because their noise is too high 
or too low. To find cosmic ray hits in the dark data, each frame up-the-ramp is 
subtracted from the subsequent frame, resulting in 199 difference frames for each ramp. 
The difference frames are then scanned to find pixels with more than 7 times
the RMS noise for this pixel. Since these are differences this is effectively a 
5 sigma effect (randomly about once per difference-frame). Then neighboring pixels are added 
to the hit if they exceed 5 times the noise. The time, position, number of 
pixels, number of electrons and other information is recorded for each hit.


\section{Results}\label{sec:results}

This section describes our characterization of cosmic ray interactions observed by NIRSpec. We begin with general considerations for hit detection, including effective detector area and energy thresholds (Section~\ref{sec:gen_considerations}). We then analyze the particle hit rates (Section~\ref{sec:rate}), the energy deposited by individual events (Section~\ref{sec:hit_energy}), their effect on noise (Section~\ref{sec:noise}), their size measured in pixels on the detectors (Section~\ref{sec:area}), and finally the nature of the largest energy deposition events (Section~\ref{sec:largest-hits}).

\subsection{General Considerations for Hit Detection}\label{sec:gen_considerations}

At the low end of the energy spectrum ($<1~\mathrm{keV}$), the imputed number of cosmic ray events is sensitive to the low energy cutoff, single-pixel event criteria, and how bad pixels are dealt with. For robustness, our reported hit characteristics and rates are therefore based on events depositing energies $>1~\mathrm{keV}$. This $1~\mathrm{keV}$ threshold intentionally excludes a population of lower-energy interactions, leading to an undercount of the total GCR event rate. Based on statistical extrapolation from the events detected above $1~\mathrm{keV}$, we estimate that these uncounted low-energy hits constitute an additional $7 \pm 2~\%$ of the particle hits reported using the $>1~\mathrm{keV}$ criterion.

Because of pixel operability differences between NIRSpec's two detectors, the detectors have slightly different effective areas for integrating cosmic ray hits. The arrays have $2048\times2048$ square pixels, with each pixel being 18~$\mu$m on a side, resulting in a geometric area of $13.59~\mathrm{cm^2}$ per array. However, the outer 4 rows and columns are reference pixels, and some science pixels exhibit high noise or are otherwise poorly functioning. The net cosmic ray detection efficiency is 98\% for NRS1 and 96.5\% for NRS2.

After correcting for this detection efficiency, the hit rates measured on the two detectors match quite well, with any residual difference (a consistently $\sim 1.78\%$ higher rate for SCA~NRS2) being within the uncertainties of conversion gain and other factors, and not statistically significant for this study. We therefore combine data from the two detectors for all subsequently reported results.

\subsection{Particle Hit Rate}\label{sec:rate}

One significant result is that NIRSpec's measured cosmic ray hit rate at L2 is somewhat less than the pre-flight prediction of $5~\mathrm{ions~cm^{-2}}~s^{-1}$ ``nominal'' \citep{Barth:2000vc}. We discuss this further in Section~\ref{sec:other-missions}, where we show that NIRSpec is likely to experience roughly $2.3-6~\mathrm{ions~cm^{-2}}~s^{-1}$ depending on solar activity, with the mean being around $4.2~\mathrm{ions~cm^{-2}}~s^{-1}$. The observed mean rate of $\sim 4.2~\mathrm{ions~cm^{-2}}~s^{-1}$ represents an approximately 16\% reduction from the pre-flight nominal prediction. While seemingly modest, this reduction translates directly into meaningful scientific benefits through its positive impact on noise.

\subsection{Particle Hit Energy}\label{sec:hit_energy}

\begin{figure}
\centering
\includegraphics[width=\columnwidth]{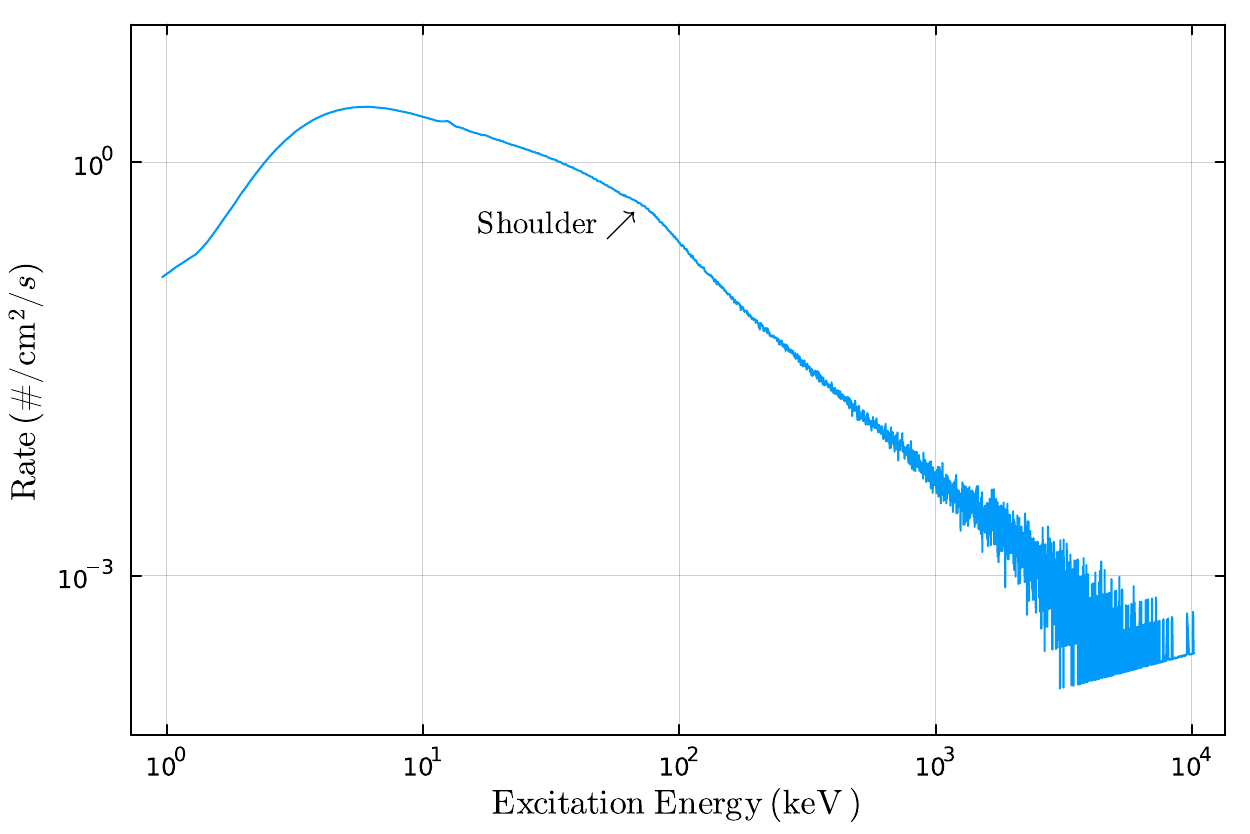}
\caption{This histogram shows the distribution of energies of the impacts. 
There is a broad peak around 6~keV which is what is expected for minimum 
ionizing protons randomly impinging on the detector. 
The shoulder around 70~keV is likely due to the 
pixel wells filling which may affect the detector efficiency for higher energy
particles. At the high end there are only a few particles in each bin which
leads to the visible shot noise. The bin width is proportional to the square
root of the energy, and then normalized so the rate is just the number per area
and time.}\label{fig:rate_vs_excitation}
\end{figure}

The observed spectrum of particle hit energies is shown in Figure~\ref{fig:rate_vs_excitation}. There is a clear peak around $6~\mathrm{keV}$, with the spectrum falling off at both lower and higher energies. This peak is consistent with the $\sim 5~\mathrm{keV}$ energy deposition predicted for normally incident minimum ionizing particles (MIP), especially when considering a range of incidence angles. The shoulder observed around $70~\mathrm{keV}$ in Figure~\ref{fig:rate_vs_excitation} corresponds to the approximate full well capacity of the pixels ($\sim60,000$~charges, or $\sim70~\mathrm{keV}$ using the average pair-production gain from Table~\ref{tab:pp-gain}). This shoulder is likely an artifact of detector response, possibly due to reduced collection efficiency as pixels saturate, rather than a true feature of the incident cosmic ray spectrum. There are a few events depositing greater than $10~\mathrm{MeV}$ total energy, but their numbers are low, and the corresponding statistics are poor. While rare, these very large events require special handling by the calibration pipeline.

To understand the expected energy deposition from MIPs, we used NIST's PSTAR calculator\footnote{\url{https://physics.nist.gov/PhysRefData/Star/Text/PSTAR.html}} to help determine elemental proton stopping powers. These are needed to calculate the energy deposition in $\mathrm{Hg_{0.7}Cd_{0.3}Te}$ from 1--3~GeV protons. Unfortunately, the PSTAR webpage does not list Hg, Cd, or Te. We therefore interpolated between the listed materials, including adjustments based on atomic number to mass ratio trends. We inferred that the mass stopping powers for 1--3~GeV protons are approximately $1.16$, $1.32$, and $1.33~\mathrm{MeV\,cm^2\,g^{-1}}$ for Hg, Cd, and Te, respectively.

We used the empirical Bragg Additivity Rule\footnote{As described in, \eg, \cite{ICRU1993_Report49}} to compute the detector's stopping power. It states that the stopping power of a compound is equal to the sum of the stopping powers of its constituent elements weighted by their mass fractions. Table~\ref{tab:detpars} shows the mass fractions for NIRSpec's detectors. Applying Bragg's rule to NIRSpec's detectors yields a combined mass stopping power of $1.25~\mathrm{MeV\,cm^2\,g^{-1}}$. This value corresponds to an energy loss of about $5~\mathrm{keV}$ for normal incidence, minimum ionizing protons traversing a $5~\mu\mathrm{m}$ thick detector layer, assuming a material density of approximately $7.87~\mathrm{g\,cm^{-3}}$.

\begin{deluxetable}{ccccccc}
    \tabletypesize{\scriptsize} 
    \tablecaption{NIRSpec $\mathrm{Hg_{1-x}Cd_xTe}$ Material Parameters}\label{tab:detpars}
    \tablehead{
    \colhead{}& \colhead{}& \colhead{Atom}& \colhead{Mass}& \colhead{Atomic}& \colhead{Atomic}& \colhead{$\mathrm{\sigma_{pn}}\tablenotemark{a}$}\\
    \colhead{Element}& \colhead{$x$}& \colhead{Fraction}& \colhead{Fraction}& \colhead{Number}& \colhead{Weight}& \colhead{(barn)}}
    \startdata
        Hg& 0.7& 0.35& 0.465& 80& 200.592& 1.8117\\
        Cd& 0.3& 0.15& 0.112& 48& 112.414& 1.2072\\
        Te& 1.0& 0.5& 0.423& 52& 127.6& 1.3154\\
    \enddata
    \tablenotetext{a}{Cross section for 1 GeV inelastic proton-nucleus scattering.}
\end{deluxetable}

\subsection{\myedit{Effect on Noise}}\label{sec:noise}

\myedit{Although Stage~1 of the \JWST pipeline finds and corrects for cosmic rays, doing so comes at a price: increased noise. This happens in two main ways. First, the pipeline can miss smaller cosmic ray hits, leaving them in the data. Second, for the cosmic rays it does find, the pipeline tries to salvage all the good data. Instead of discarding the ramp after a hit, it calculates the slope of the ramp before and after the hit when possible. It then computes a weighted average of these two slopes to get the final count rate. The trade-off is that each of those shorter segments gives a less certain—and therefore noisier—slope measurement. Averaging them results in a final rate that is still noisier than what one would get from a single, uninterrupted integration.}

Before launch, \citet{Giardino2019} used Monte Carlo simulation to predict that cosmic rays would degrade NIRSpec's total noise by about 7\% compared to the on-ground performance. \citet{Birkmann2022} subsequently reported that this prediction had borne out after launch. Knowing the average number of pixels disturbed by cosmic rays is therefore an important parameter for Exposure Time Calculators (ETC).

\subsection{Particle Hit Area}\label{sec:area}

\begin{figure}[!htbp]
  \centering
  \includegraphics[width=\columnwidth]{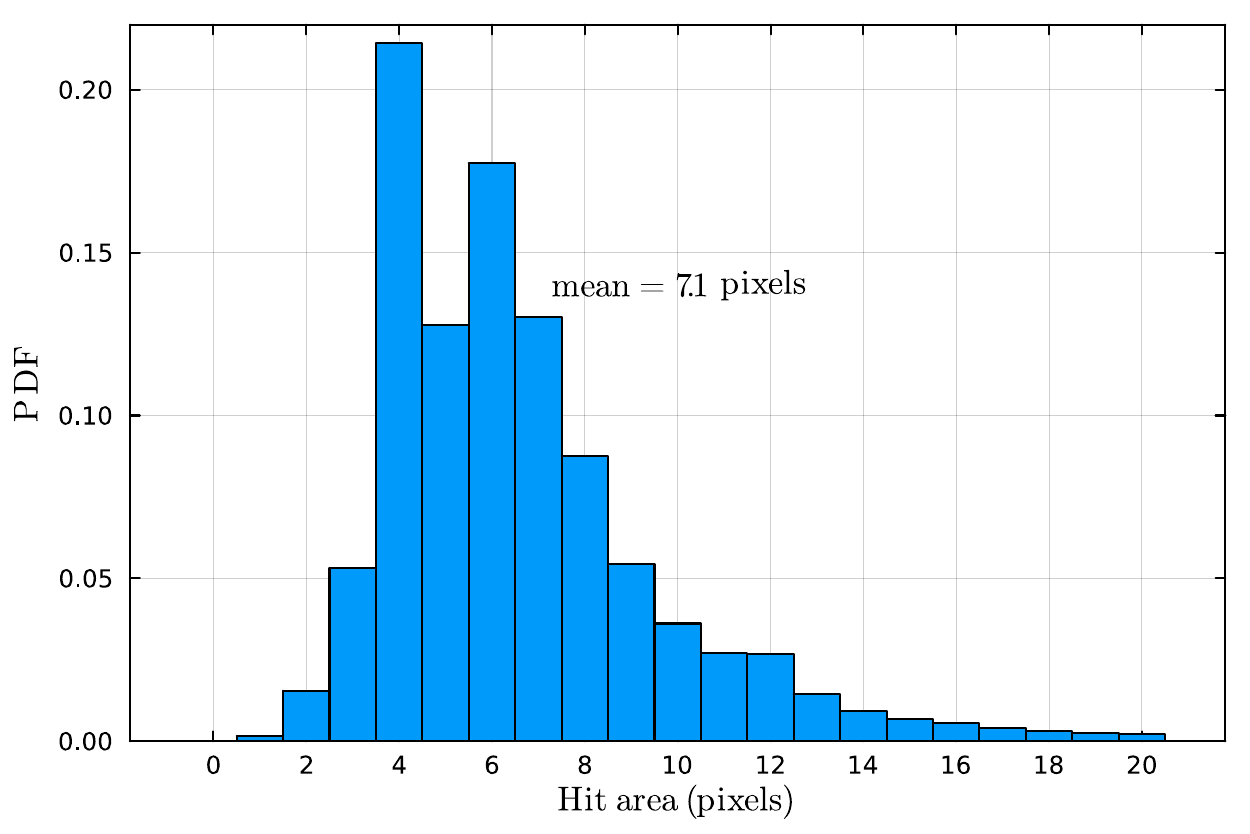}
  \caption{This histogram shows the probability density function of cosmic ray hit sizes. It combines data from the two NIRSpec detectors. The mean hit area is 7.1~pixels. The local minimum at 5~pixels is a robust feature, appearing in both detectors and in different subsets of data. Similar minima can be reproduced in Monte Carlo simulations using the measured hit rate versus energy spectrum (see Figure~\ref{fig:rate_vs_excitation}) as input. In these simulations, the precise characteristics of the local minimum (such as its depth, exact position, or number of local minima) were found to be sensitive to the selection thresholds applied. \myedit{In general, hit area correlates with energy deposition. We have cropped off some very energetic hits to show the main histogram better.}}\label{fig:area_histo}
\end{figure}

Figure~\ref{fig:area_histo} shows the measured cosmic ray hit area histogram. The mean hit area is about 7.1~pixels. This observed mean is smaller than the typical hit size of up to 9 pixels assumed in pre-flight \JWST ETC planning.

We were initially surprised by the local minimum  at 5 pixels. It is a robust feature that appears in both NIRSpec detectors, in multiple cuts through the data, and in other \JWST instruments \citep{Martel2024}. Intuitively, we had expected the peak to be at 5~pixels because a perfectly centered hit at normal incidence would clearly disturb the hit pixel. We furthermore expected inter-pixel capacitance \citep[IPC;][]{Moore2006} to couple it to the four nearest neighbors yielding an initial estimate of 5~pixels.

To further explore this, we constructed simple Monte Carlo models. We found that the shape of the energy spectrum (Figure~\ref{fig:rate_vs_excitation}) produces a local minimum at 5~pixels, and that we could change the detailed shape of the histogram by varying the hit detection thresholds. We therefore believe that the local minimum at 5~pixels is caused by the shape of the cosmic ray energy spectrum and tunable by adjusting detection thresholds.

\subsection{The Largest Cosmic Ray Events}\label{sec:largest-hits}

The largest observed cosmic ray events often manifest as ``snowballs''; or more rarely, they appear as extremely intense events with associated showers.\footnote{\url{https://jwst-docs.stsci.edu/known-issues-with-jwst-data/shower-and-snowball-artifacts\#ShowerandSnowballArtifacts-Snowballssnowballs}}\label{foot:showers} The term snowball was coined circa 2005 to describe very large, circular, saturating, cosmic-ray-like signatures identified during ground testing of \JWST and \HST WFC3 detectors.

On orbit, NIRSpec sees something similar to the historical snowballs, as well as more exotic events including some with showers of secondary particles. The showers that we see in NIRSpec data are different from the shower phenomenon that has been described in the \JWST MIRI.\footref{foot:showers}

The definitive root cause of snowballs observed in ground tests was not established before \JWST's launch, though alpha-emitting radioactive contaminants on or near the detectors were considered a possible source. On orbit, the large energy depositions, which manifest as both compact snowballs and, more rarely, as energetic events with associated secondary showers, are thought to originate from several mechanisms.

One candidate is the direct impact of highly ionizing GCRs, such as heavy ions; an iron nucleus, for example, deposits substantially more energy ($Z^2 \approx 676$ times greater for charge $Z=26$) than a proton of the same velocity and can readily cause extensive pixel saturation. Alternatively, these large events might result from secondary particles produced when primary GCRs interact within the observatory's shielding or the detector material itself. Among such secondaries, those with low energy (e.g., alpha particles or nuclear spallation fragments) that stop within the thin detector layer could explain snowballs that appear almost perfectly circular. In contrast, events that are elongated and feature collimated showers of particles may be more consistent with complex interactions involving surrounding material and the detector material, or occasionally, high-energy inelastic scattering of GCRs occurring directly within the detector material.

\begin{figure*}
\centering
\includegraphics[width=\textwidth]{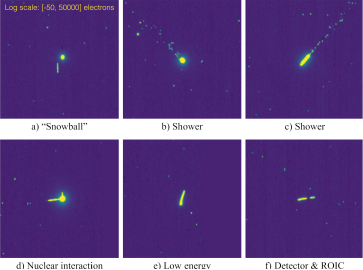}
\caption{Examples of large cosmic ray hit morphologies. (a) Snowballs are round, with saturated cores, and usually no showers. (b-c) Secondary showers. (d) Possible inelastic scattering. (e) Curved track potentially from low-energy particle. (f) Broken track possibly due to particle traversal of both detector material and ROIC.}\label{fig:cr_examples}
\end{figure*}

Figure~\ref{fig:cr_examples} presents six examples of very large cosmic ray hits observed on the NIRSpec detectors.
Panel (a) of Figure~\ref{fig:cr_examples} displays a roughly circular, strongly saturated hit, often referred to as a snowball.
Panels (b) and (c) exhibit more complex structures identified as secondary showers.
These features are consistent with high-energy GCRs undergoing complex interactions with the detector and surrounding material or perhaps nuclear interactions within the HgCdTe detector material, leading to an often collimated cascade of secondary particles.
A different type of nuclear interaction is depicted in panel (d), characterized by a significant deviation in the primary particle's path or the creation of a short, highly ionizing recoil track.
Panel (e) illustrates a curved track, which suggests the passage of a very low-energy particle; such a trajectory could be produced by, for instance, a secondary alpha particle originating from radioactive decay within the instrument shielding or nearby materials.
Finally, panel (f) shows a broken or discontinuous track. This morphology might be attributed to a particle traversing both the HgCdTe detector layer and the underlying Readout Integrated Circuit (ROIC). These two components are separated by a gap of approximately a few microns, and a particle crossing this void would not deposit charge, leading to the observed break in the track.

\begin{figure}
\centering
\includegraphics[width=\columnwidth]{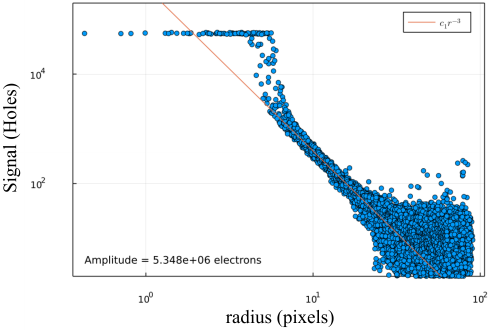}
\caption{Radial profile of a typical large snowball halo cosmic ray event. The plot shows the saturated core, the shoulder region with a steep decline in charge, and the extended halo exhibiting an $r^{-3}$ falloff (indicated by the fitted line). The x-axis represents radial distance from the core, and the y-axis represents signal intensity or charge.}
\label{fig:snowball_halo}
\end{figure}

Snowballs have a typical structure (Figure~\ref{fig:snowball_halo}). At the center is a heavily saturated core, where the charge deposited by the particle exceeds the full well capacity of the pixels. Surrounding this core is a ``shoulder'' region, where the collected charge diminishes rapidly with increasing radial distance from the impact center. Beyond this shoulder extends a faint, diffuse halo. A key characteristic of this extended halo is that its surface brightness profile typically follows an $\sim r^{-3}$ power law, where $r$ is the radial distance from the event's core. We do not have a satisfactory explanation for this empirical relation at this time.

Some of the hits like those shown in Figure~\ref{fig:cr_examples}(b-d) suggest inelastic scattering in the HgCdTe detector material. To test this hypothesis, we computed the probability that a 1~GeV proton might undergo inelastic scattering while passing through the $\sim5~\mu$m thick detector layer at $45^\circ$ incidence.

\citet{Letaw1983} provides an empirical formula for the proton-nucleus inelastic scattering cross sections of Hg, Cd, and Te that is accurate to a few percent at 1~GeV,
\begin{multline}
\sigma\left(E\right) = 45 A^{0.7}\left[1 + 0.016 \sin\left(5.3-2.63\ln A\right) \right] \times\\
\left[1 - 0.62e^{-E/200}\sin\left(10.9E^{-0.28}\right)\right]~\mathrm{mb}.\label{eq:cross_section}
\end{multline}
In this expression, $A$ is atomic mass and $E$ is proton energy in MeV. Table~\ref{tab:detpars} evaluates this expression for NIRSpec's detectors. To compute the total inelastic scattering cross section per unit volume, one sums the cross sections of the constituent atoms. This is straightforwardly done using the density and mass fractions of the elements from Table~\ref{tab:detpars}.

Assuming the same $\rho=7.87~\mathrm{g~cm^{-3}}$ density as before, the inelastic scattering cross section per unit volume for 1~GeV protons in NIRSpec's detectors is $\mu=4.63\times10^{-6}\mu\mathrm{m}^{-1}$. The probability of any one proton, incident at $45^\circ$, undergoing inelastic scattering is,
\begin{equation}
p = 1 - e^{-\mu\sqrt{2}\ell},
\end{equation}
where $\ell$ is the HgCdTe detector layer's thickness. With $\ell=5~\mu$m, the result is $p=3.27\times10^{-5}$.

The average GCR rate since launch has been about $3.3~\mathrm{ions~cm^{-2}}~s^{-1}$. Integrating over the entire data set and assuming that 90\% of the events are protons, about $3.6\times10^7$ protons struck NIRSpec's detectors, and of these we would expect a little over 1,000 to have undergone inelastic scattering. While we have not tried to carefully count showers and other evidence of potential inelastic scattering, this is broadly consistent with the observation that showers are a very rare yet nevertheless real feature in the data. Based on this, inelastic scattering in the HgCdTe seems a plausible explanation for at least some events like those in Figure~\ref{fig:cr_examples}(b-d).

Although events with large haloes constitute only a small fraction of GCRs, they are nevertheless important to the pipeline. Snowballs were first identified because of their round, completely saturated cores. Arbitrarily setting 9 pixels as the lower limit on core size, an event must dissipate $\gtrsim 1$~MeV in the detector to potentially create a snowball. In these data, the corresponding snowball rate is about 0.3 snowballs per 14.58889 second NIRSpec frame, or $0.0015~\mathrm{snowballs~cm^{-2}}~s^{-1}$. \myedit{This is very roughly $>100\times$ higher than was observed during \JWST ground integration and testing.}\footnote{\protect\myedit{Although we see snowballs both on the ground at at L2, the vastly different rates indicate different causative mechanisms.}} Any full-frame NIRSpec integration longer than about a minute will probably have some snowballs in it. The pipeline therefore needs to be aware that pixels potentially far from a snowball might be disturbed and treat them accordingly.

Providing detailed explanations for the complex, high energy events is beyond the scope of this paper. In September, 2023, the authors gave a seminar at Fermilab. Many attendees were particle physicists. Their consensus view was that the high energy events could probably be explained using Geant4 (GEometry ANd Tracking 4)\footnote{\url{https://geant4.web.cern.ch/}}. \myedit{We would be happy to share our catalog of NIRSpec cosmic ray hits upon request.}

\section{Comparison to Other Missions}\label{sec:other-missions}

\begin{figure}[!htbp]
  \centering
  \includegraphics[width=\columnwidth]{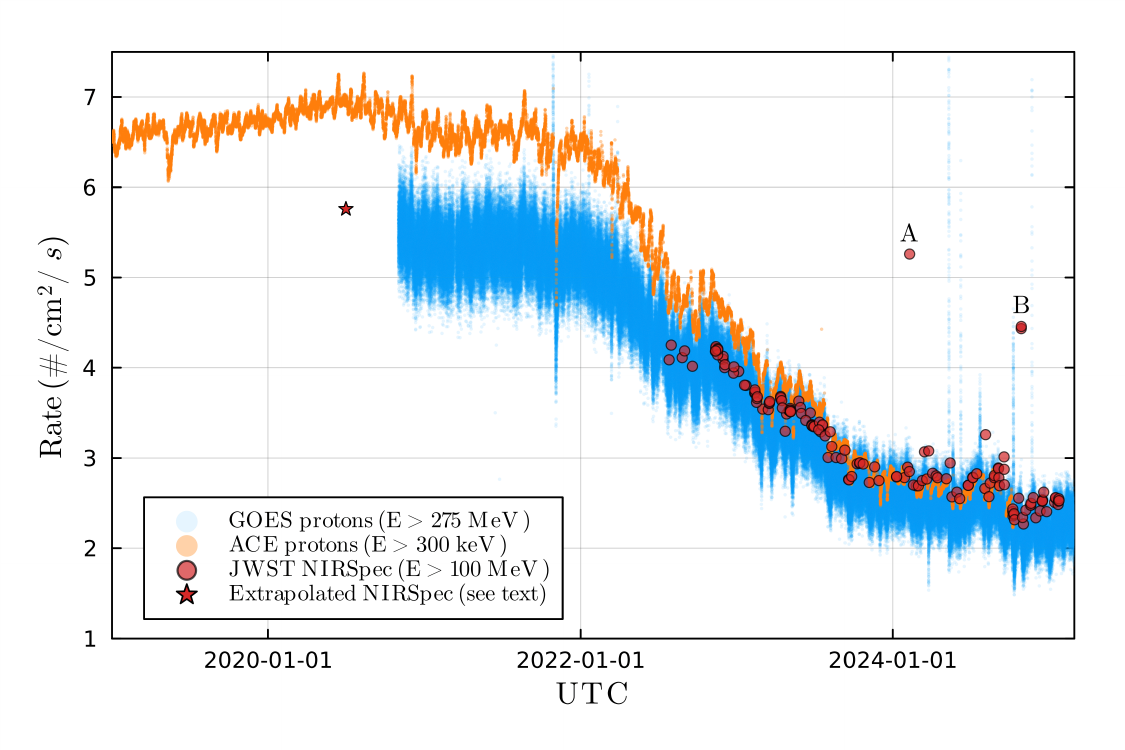}
  \caption{This figure shows the cosmic ion fluxes seen by NIRSpec's two detectors  overlaid on some cosmic proton fluxes seen by relevant missions. \ACE is located in deep space at L1. The ACE plot is fairly smooth because the downloadable data exclude periods of high solar activity. For comparison, we show GOES-16. As described in the text, we have integrated the GOES proton data over energies $\mathrm{E>100~MeV}$ to mimic the effects of NIRSpec's shielding. The GOES flux is lower because the Earth's magnetosphere provides additional shielding of GCRs in GOES' geostationary orbit that \JWST does not experience at L2. NIRSpec happened to be acquiring calibration darks during two periods of enhanced solar activity. These are labeled A and B. This shows that some solar particles are able to get through during periods of enhanced solar activity, although the effect would not have stopped science observations during these periods. \JWST launched near solar minimum, and it has seen nearly the full solar cycle. During this period, cosmic ray fluxes ranged from about $4.3-2.3~\mathrm{ions~cm^{-2}}~s^{-1}$ excluding points A and B. During the previous solar cycle, \citet{Kirsch2018} measured a range of $2-4~\mathrm{ions~cm^{-2}}~s^{-1}$ using Gaia.}\label{fig:rates}
\end{figure}

Figure~\ref{fig:rates} shows the measured \JWST
NIRSpec hit rates overlaid on hit rates from two relevant space missions, ACE and GOES. \myedit{This section provides more background on \ACE and \GOES.} Then, building on all available data, we discuss what \JWST and \RomanST should see in the next few years.

\subsection{ACE}\label{sec:ace}

NASA's \ACE was launched into a Sun/Earth-Moon L1 Lissajous orbit in August, 1997 \citep{Stone1998SSRv...86....1S}. This orbit was chosen to provide continuous monitoring of the solar wind, interplanetary magnetic field, and energetic particle populations before they interact with the Earth's magnetosphere. ACE's primary scientific objective is to determine and compare the elemental and isotopic composition of matter from diverse sources, including the solar corona (via solar wind and solar energetic particles, SEPs), the interplanetary medium, the local interstellar medium, and GCRs \citep{Chiu1998SSRv...86..257C}. During quiescent periods, it is mostly high energy GCRs that have sufficient energy to penetrate NIRSpec's radiation shielding.

ACE characterizes the space radiation environment using a suite of nine instruments \citep{Stone1998SSRv...86....1S}. Of these, the Cosmic Ray Isotope Spectrometer (CRIS) is particularly important for characterizing GCRs relevant to this study. CRIS is designed to measure the elemental and isotopic composition of GCR nuclei from Helium, $\mathrm{Z=2}$ to Zinc $\mathrm{Z=30}$ ``and beyond'',\footnote{\url{https://izw1.caltech.edu/ACE/CRIS_SIS/cris.html\#gcrc}} using four identical stacks of large-area silicon solid-state detectors. Each stack includes detectors designated E1 through E8 for energy deposition measurements. An additional detector, E9, located at the bottom of each stack, is used in anticoincidence to identify and guard against particles that penetrate the entire detector stack. The ``singles rate'' from this E9 detector provides a measure of the flux of these highly penetrating particles.

Although the overall radiation environment at L1 differs from that at L2, the most important differences have to do with plasmas that are mostly blocked by \JWST structure and the NIRSpec detectors' radiation shielding. Since both L1 and L2 are well outside Earth's magnetosphere, we expect the GCR environments to be similar for ACE and \JWST.

For Figure~\ref{fig:rates}, we downloaded  contributed ``Level 3'' data from the Ace Science Center.\footnote{\url{https://izw1.caltech.edu/ACE/ASC/DATA/level3/cris/GCRprotons/CRIShourlyE9rates.txt}} According to the website, these are the hourly averaged CRIS E9 rate, after cuts to clean the data of artifacts and remove active periods (SEPs). According to a note provided with the data, during solar quiet periods the
singles count rate is probably dominated by GCR protons at energies $>120~\mathrm{MeV}$. The note cautions that the singles data are not carefully calibrated. But, because of the detector's large area (4 10-cm diameter 3-mm thick Si detectors) the statistical accuracy of its count rate is quite high. For our purposes, we treat the ACE data as one of several points of comparison for the NIRSpec data that are our main focus.

\subsection{GOES-16}\label{sec:goes}

\GOES-16 (hereafter ``\GOES'') directly measures proton fluxes in the outer regions of the Earth's magnetosphere. Although GOES's GEO orbit is inside the Earth's magnetosphere, the magnetic field is quite weak at $r_\textrm{GEO}=6.6~r_\earth$. One would 
therefore expect the Earth's magnetosphere to provide only weak shielding of cosmic rays sufficiently energetic to penetrate NIRSpec's shielding in GEO.

The most relevant instrument for our purposes is the Space Environment In-Situ Suite (SEISS) Solar and Galactic Proton Sensor (SGPS). It covers a broad energy range from 1 MeV to $>$500 MeV. The SGPS differential channels were designed and calibrated to measure Solar Particle Events (SPE). These are characterized by high particle flux that falls
off rapidly with increasing particle energy.  Most SPE particles are blocked
in NIRSpec's shielding. However, high energy GCRs pass completely through. Fortunately, SPGS channels P10 (275-500 MeV) and P11 ($> 500~\mathrm{MeV}$) are calibrated for use during solar quiet conditions.

\subsection{Predictions for \JWST}

The Sun modulates GCRs in the solar system. The GCR rate is low when the Sun is active and vice versa. The mechanism is thought to be the stronger, tangled magnetic fields that appear in the heliosphere when the sun is particularly active. These act to deflect charge particles away from the inner solar system.

\begin{figure}
\centering
\includegraphics[width=\columnwidth]{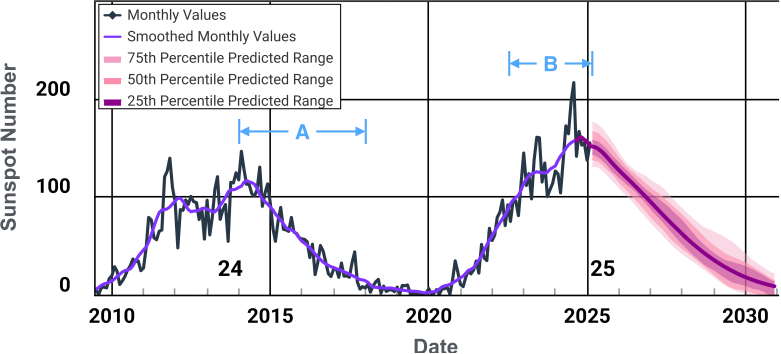}
\caption{Sunspot number is correlated with solar activity and inversely correlated with the GCR rate. This plot shows sunspot number during: (A) a period when the Gaia CCDs saw between $2-4~ \mathrm{protons~cm^{-2}}~s^{-1}$ \citep{Kirsch2018} and (B) this study. Like \JWST, Gaia is in an L2 orbit. Credit: National Solar Observatory, \url{https://www.swpc.noaa.gov/products/solar-cycle-progression}}\label{fig:solar_cycle}
\end{figure}

Figure~\ref{fig:solar_cycle} plots the number of sunspots versus date during the current and last solar cycle. It shows that at the time of \JWST's launch in December, 2021, solar activity was near its minimum. The first 6~months of \JWST's mission were spent commissioning the observatory. \JWST started collecting science data, and taking regular calibraiton darks, in June, 2022. This is indicated by region B in the plot.

Looking forward, we expect \JWST to experience the declining phase of Solar Cycle 25. As solar activity falls, the cosmic ray rate will increase from about $2.3-4.3~\mathrm{ions~cm^{-2}}~s^{-1}$ between now and mid-2027. It will then continue to increase, peaking perhaps around $6~\mathrm{ions~cm^{-2}}~s^{-1}$ in the early 2030s. The cycle will then repeat itself scaled by the strength of the next Solar Cycle.

We see no reason why the characteristics of individual hits would change significantly. We therefore expect the typical hit to disturb about 7.1 pixels sufficiently to require mitigation. 

\section{Expectations for Roman Space Telescope}\label{sec:roman}

\RomanST, scheduled for launch no later than May, 2027, is NASA's next astrophysics flagship mission. Similar to \JWST, \RomanST will operate from a L2 quasi-halo orbit. This will expose it to a GCR environment similar to \JWST's. \RomanST's launch is anticipated during the declining phase of Solar Cycle 25, roughly midway between its recent maximum and the next minimum (see Figure~\ref{fig:solar_cycle}).

For its Wide Field Imager (WFI), \RomanST will use Teledyne H4RG-10 HgCdTe detectors with a shorter cutoff wavelength ($\lambda_\textrm{co}=2.5~\mu$m) than NIRSpec's. \citet{Mosby2020} describes these detectors in detail. Key detector differences from NIRSpec will influence cosmic ray manifestations:

\begin{enumerate}
\item \textbf{Smaller Pixels:} $10~\mu$m pitch versus NIRSpec's $18~\mu$m.
\item \textbf{Thinner Active Layer:} The HgCdTe layer is likely a little thinner.
\item \textbf{Reduced Charge Diffusion:} Lateral charge spread is anticipated to be less on account of the somewhat thinner active layer and higher photodiode reverse bias.
\end{enumerate}

The WFI's radiation shielding's effectiveness is probably similar to NIRSpec's. The H4RG-10's molybdenum package is broadly similar to the NIRSpec H2RG package vis-\`a-vis shielding. 

The detector differences will affect how hits manifest. While a thinner layer and less diffusion might imply smaller charge clouds, the significantly smaller pixels might cause an event to disturb more pixels. A typical \RomanST hit might affect somewhat more pixels than NIRSpec's $\sim 7.1$, though the increase should not be drastic.

\RomanST's $\lambda_\textrm{co}=2.5~\mu$m cutoff implies that its HgCdTe has a bandgap roughly twice as wide as that of NIRSpec's $\lambda_\textrm{co}=5.3~\mu$m material. This wider bandgap will halve the number of electron-hole pairs generated per keV of deposited energy. A typical \RomanST GCR event will likely produce only about half the charge in the detector layer as a similar NIRSpec event.

Regarding GCR interaction rates, assuming solar activity (the remainder of Solar Cycle 25 and the onset of Solar Cycle 26) follows typical modulation patterns, \RomanST will likely launch into a GCR flux of approximately $4.3~\mathrm{ions~cm^{-2}}~s^{-1}$. Towards the Solar Cycle 25 minimum (expected in the early 2030s), this rate could increase to $\sim 6~\mathrm{ions~cm^{-2}}~s^{-1}$. Subsequently, as Solar Cycle 26 activity rises, the GCR flux should decrease, possibly returning to $4.3~\mathrm{ions~cm^{-2}}~s^{-1}$ roughly five years post-launch (circa early 2032).

\section{Summary}\label{sec:summary}

We describe cosmic ray impacts on \JWST NIRSpec's $\lambda_\textrm{co}=5.4~\mu$m cutoff HgCdTe H2RG detectors using dark exposures from NIRSpec's first three years of operations at L2. These detectors do not directly see the L2 radiation environment. Rather, they sit behind shielding that was designed to meet a requirement of greater than 10~mm aluminum-equivalent over $4\pi$ steradians. The shielding effectively blocks solar particles except during periods of high solar activity. High energy GCRs ($\mathrm{E>100~keV}$) penetrate the shielding and dominate the observed events. Our findings potentially inform ongoing \JWST operations and future missions at L2 and in deep space having comparable shielding.

The NIRSpec GCR hit rate decreased from about 4.3 to $2.3~\mathrm{ions~cm^{-2}}~s^{-1}$ during the period June, 2022, to February, 2025. This was driven by increasing solar activity on the rising side of Solar Cycle 25. A typical GCR hit affects $\sim 7.1$ pixels and deposits $\sim 6~\mathrm{keV}$ (equivalent to $\sim 5200$ charges) in the detector material. This corresponds to a linear energy transfer of $\sim 0.86~\mathrm{keV~\mu m^{-1}}$. These characteristics overwhelm the read noise and highlight the need for robust cosmic ray mitigation, as is done in Stage 1 of the \JWST pipeline.

Our investigation explored a surprisingly high number of large snowball events that were seen on-orbit (Section~\ref{sec:largest-hits}). Although snowballs were seen during ground testing, we see very roughly $100\times$ more of them on-orbit. Some of the largest events have associated secondary showers.

\myedit{Very large cosmic ray hits may originate from several sources: direct impacts from heavy ions, secondary particles produced within NIRSpec's shielding, or complex interactions within or near the detector material, such as inelastic scattering within the HgCdTe, that produce particle showers.}

Looking ahead for \JWST, we expect the GCR hit rate to rise as solar activity declines. The rate should increase from its current $\sim 2.3~\mathrm{ions~cm^{-2}}~s^{-1}$ to $\sim 4.3~\mathrm{ions~cm^{-2}}~s^{-1}$ by early 2027, potentially reaching $\sim 6~\mathrm{ions~cm^{-2}}~s^{-1}$ near the Solar Cycle 25 minimum in the early 2030s. Because the snowball rate tracks the GCR rate, we expect to see snowballs following the same trend.

These results should apply to \RomanST (Section~\ref{sec:roman}). Planned for launch no later than May, 2027, \RomanST will enter its L2 orbit as Solar Cycle 25 wanes. \RomanST will likely experience an initial GCR flux near $4.3~\mathrm{ions~cm^{-2}}~s^{-1}$. This rate is projected to rise to $\sim 6~\mathrm{ions~cm^{-2}}~s^{-1}$ during solar minimum before declining back to $\sim 4.3~\mathrm{ions~cm^{-2}}~s^{-1}$ near \RomanST's fifth year at L2.

While \RomanST's radiation shielding effectiveness should be comparable to NIRSpec's, its H4RG-10 HgCdTe detectors ($\lambda_\mathrm{co} \approx 2.5~\mu\mathrm{m}$) differ significantly. H4RG-10 detectors feature smaller pixels ($10~\mu\mathrm{m}$ versus $18~\mu\mathrm{m}$), perhaps a somewhat thinner absorber layer, and reduced charge diffusion on account of the thinner active layer. Though \RomanST's hits may affect a somewhat larger number of its smaller pixels, we do not expect a dramatic difference. However, \RomanST's wider detector bandgap will halve the electron-hole pairs generated per keV dissipated. A \RomanST GCR hit will produce only about half the charge seen in a NIRSpec event.

In summary, we find that cosmic rays at L2 are somewhat less disruptive for \JWST NIRSpec than pre-launch expectations. Before launch, \JWST planned for $5~\mathrm{ions~cm^{-2}}~s^{-1}$ ``nominal'', and for each event to disturb potentially as many as nine pixels. In practice, NIRSpec's average observed flux, if extrapolated over a full solar cycle, is about 16\% lower than the pre-launch ``nominal'' expectation and typical events are about 20\% smaller in pixels. Although very large snowball events, and events with showers are comparatively rare, they are nevertheless present in most full frame exposures. The largest events present unique challenges to calibration pipelines and would benefit from further study, \myedit{perhaps using Geant4 to more fully understand the physical mechanisms.}

\begin{acknowledgments}
This work was supported by NASA as part of the James Webb and Nancy Grace Roman Space Telescope Projects. Resources supporting this work were provided by the NASA High-End Computing (HEC) Program through the NASA Center for Climate Simulation (NCCS) at Goddard Space Flight Center. We thank Maurice te Plate for helpful discussions about NIRSpec's radiation shielding. We acknowledge the use of Google Gemini for assistance with spelling/typo correction, grammar, and stylistic improvements to this manuscript.
\end{acknowledgments}

%

\vspace{5mm}
\facilities{\JWST NIRSpec}


\software{Julia \citep{Bezanson2017}}






\bibliography{ms}{}

@article{Alig1977,
author = {Alig, R C and Bloom, S},
number = {11},
pages = {677--680},
journal = {Phys. Rev. Lett.},
title = {{Electron-Hole-Pair Creation Energies in Semiconductors}},
volume = {35},
year = {1977}
}

@article{Barth:2000vc,
  author = {Barth, Janet L and Isaacs, John C and Poivey, Christian},
  title = {{The Radiation Environment for the Next Generation Space Telescope}},
  journal={Next Generation Space Telescope Program},
  year = {2000}
}

@article{Birkmann2022,
author = {Birkmann, Stephan M. and Giardino, Giovanna and Sirianni, Marco and Ferruit, Pierre and Rauscher, Bernhard J. and {Alves de Oliveira}, Catarina and B{\"{o}}ker, Torsten and Kumari, Nimisha and L{\"{u}}tzgendorf, Nora and Manjavacas, Elena and Proffitt, Charles and Rawle, Tim and te Plate, Maurice and Zeidler, Peter},
journal = {Proc SPIE},
pages = {101},
title = {{The in-flight noise performance of the JWST/NIRSpec detector system}},
volume = {12180},
year = {2022}
}

@article{Bezanson2017,
author = {Bezanson, Jeff and Edelman, Alan and Karpinski, Stefan and Shah, Viral B},
doi = {10.1137/141000671},
journal = {SIAM Rev.},
number = {1},
pages = {65--98},
title = {{Julia: A Fresh Approach to Numerical Computing}},
url = {https://doi.org/10.1137/141000671},
volume = {59},
year = {2017}
}

@article{Chiu1998SSRv...86..257C,
    author         = {{Chiu}, M.~C. and {Haggerty}, D.~K. and {Bustard}, W.~C. and {Burley}, R.~J. and {Deckert}, K.~K. and {Dennis}, K.~L. and {Fountain}, G.~H. and {Hofmann}, D.~J. and {Hunt}, Jr., J.~W. and {Kennel}, H.~F. and {Krueger}, W.~F. and {Ross}, M.~I. and {Stadter}, J.~T. and {Williams}, R.~L.},
    title          = "{ACE Spacecraft}",
    journal        = {Space Science Reviews},
    year           = {1998},
    month          = {Jun},
    volume         = {86},
    number         = {1-4},
    pages          = {257-284},
    doi            = {10.1023/A:100501022ACE},
    adsurl         = {https://ui.adsabs.harvard.edu/abs/1998SSRv...86..257C}
}

@article{Evans2003,
  title={Natural environment near the Sun/Earth-Moon L2 libration point},
  author={Evans, Steven W and others},
  journal={Next Generation Space Telescope Program},
  year={2003},
  publisher={NASA Greenbelt, MD, USA}
}

@article{Fox2009,
author = {Fox, Ori and Waczynski, Augustyn and Wen, Yiting and Foltz, Roger D and Hill, Robert J and Kimble, Randy A and Malumuth, Eliot and Rauscher, Bernard J},
doi = {10.1086/605131},
journal = {PASP},
month = {jul},
number = {881},
pages = {743--754},
publisher = {{\{}IOP{\}} Publishing},
title = {{The55Fe X-Ray Energy Response of Mercury Cadmium Telluride Near-Infrared Detector Arrays}},
url = {https://doi.org/10.1086{\%}2F605131},
volume = {121},
year = {2009}
}

@article{Giardino2019,
author = {Giardino, Giovanna and Birkmann, Stephan and Robberto, Massimo and Ferruit, Pierre and Rauscher, Bernard J and Sirianni, Marco and Oliveira, Catarina Alves De and Boeker, Torsten and Luetzgendorf, Nora and Plate, Maurice and Puga, Elena and Rawle, Tim},
doi = {10.1088/1538-3873/ab2fd6},
issn = {1538-3873},
journal = {PASP},
number = {1003},
pages = {94503},
publisher = {IOP Publishing},
title = {{The Impact of Cosmic Rays on the Sensitivity of JWST / NIRSpec}},
url = {http://dx.doi.org/10.1088/1538-3873/ab2fd6},
volume = {131},
year = {2019}
}

@book{ICRU1993_Report49,
  author    = {{International Commission on Radiation Units and Measurements (ICRU)}},
  title     = {Stopping Powers and Ranges for Protons and Alpha Particles},
  series    = {ICRU Report},
  number    = {49},
  publisher = {{International Commission on Radiation Units and Measurements}},
  year      = {1993},
  address   = {Bethesda, MD}
}

@article{Jakobsen_2022,
author = {Jakobsen, P and Ferruit, P and {Alves de Oliveira}, C and Arribas, S and Bagnasco, G and Barho, R and Beck, T.{\~{}}L. and Birkmann, S and B{\"{o}}ker, T and Bunker, A.{\~{}}J. and Charlot, S and de Jong, P and de Marchi, G and Ehrenwinkler, R and Falcolini, M and Fels, R and Franx, M and Franz, D and Funke, M and Giardino, G and Gnata, X and Holota, W and Honnen, K and Jensen, P.{\~{}}L. and Jentsch, M and Johnson, T and Jollet, D and Karl, H and Kling, G and K{\"{o}}hler, J and Kolm, M -G. and Kumari, N and Lander, M.{\~{}}E. and Lemke, R and L{\'{o}}pez-Caniego, M and L{\"{u}}tzgendorf, N and Maiolino, R and Manjavacas, E and Marston, A and Maschmann, M and Maurer, R and Messerschmidt, B and Moseley, S.{\~{}}H. and Mosner, P and Mott, D.{\~{}}B. and Muzerolle, J and Pirzkal, N and Pittet, J -F. and Plitzke, A and Posselt, W and Rapp, B and Rauscher, B.{\~{}}J. and Rawle, T and Rix, H -W. and R{\"{o}}del, A and Rumler, P and Sabbi, E and Salvignol, J -C. and Schmid, T and Sirianni, M and Smith, C and Strada, P and te Plate, M and Valenti, J and Wettemann, T and Wiehe, T and Wiesmayer, M and Willott, C.{\~{}}J. and Wright, R and Zeidler, P and Zincke, C},
doi = {10.1051/0004-6361/202142663},
journal = {A{\&}A},
month = {may},
pages = {A80},
title = {{The Near-Infrared Spectrograph (NIRSpec) on the James Webb Space Telescope. I. Overview of the instrument and its capabilities}},
volume = {661},
year = {2022}
}

@article{Kirsch2018,
author = {Kirsch, Christian Thomas},
journal = {Master's Thesis Physics, Dr. Karl Remeis-Sternwarte Astron. Inst.},
title = {{The Cosmic Ray Background at L2 as Seen in Gaia Observations}},
year = {2018}
}

@article{Letaw1983,
author = {Letaw, J.{\~{}}R. and Silberberg, R and Tsao, C.{\~{}}H.},
doi = {10.1086/190849},
journal = {ApJ Supp.},
month = {mar},
pages = {271--275},
title = {{Proton-nucleus total inelastic cross sections - an empirical formula for E greater than 10 MeV}},
volume = {51},
year = {1983}
}

@article{Loose:2003vh,
author = {Loose, Markus and Farris, Mark C and Garnett, James D and Hall, Donald N B and Kozlowski, Lester J},
journal = {Proc SPIE},
month = {mar},
pages = {867},
title = {{HAWAII-2RG: a 2k x 2k CMOS multiplexer for low and high background astronomy applications}},
volume = {4850},
year = {2003}
}

@inproceedings{Loose2007,
author = {Loose, Markus and Beletic, James and Garnett, James and Xu, Min},
booktitle = {Proc SPIE},
doi = {10.1117/12.735625},
editor = {Grycewicz, Thomas J and Marshall, Cheryl J and Warren, Penny G},
month = {sep},
pages = {66900C},
series = {Society of Photo-Optical Instrumentation Engineers (SPIE) Conference Series},
title = {{High-performance focal plane arrays based on the HAWAII-2RG/4G and the SIDECAR ASIC}},
volume = {6690},
year = {2007}
}

@article{Martel2024,
author = {Martel, A.R. and Cooper, R. and Volk, K.},
journal = {JWST Tech. Rep.},
title = {{The Population of Cosmic Rays and Snowballs Detected in the JWST NIRISS Instrument}},
volume = {JWST-STScI-008875},
year = {2024}
}

@article{Moore2006,
author = {Moore, Andrew C.},
doi = {10.1117/1.2219103},
issn = {0091-3286},
journal = {Opt. Eng.},
number = {7},
pages = {076402},
title = {{Quantum efficiency overestimation and deterministic cross talk resulting from interpixel capacitance}},
volume = {45},
year = {2006}
}

@article{Mosby2020,
author = {Mosby, Gregory and Rauscher, Bernard J and Bennett, Chris and Cheng, Edward S and Cheung, Stephanie and Cillis, Analia and Content, David and Cottingham, Dave and Foltz, Roger and Gygax, John and Hill, Robert J and Kruk, Jeffrey W and Mah, Jon and Meier, Lane and Merchant, Chris and Miko, Laddawan and Piquette, Eric C and Waczynski, Augustyn and Wen, Yiting},
doi = {10.1117/1.JATIS.6.4.046001},
journal = {JATIS},
month = {oct},
pages = {46001},
title = {{Properties and characteristics of the Nancy Grace Roman Space Telescope H4RG-10 detectors}},
volume = {6},
year = {2020}
}

@article{Pickel2002,
author = {Pickel, J. C. and Reed, Robert A. and Ladbury, R. and Rauscher, B. and Marshall, Paul W. and Jordan, Tom M. and Fodness, B. and Gee, G.},
doi = {10.1109/TNS.2002.805382},
issn = {00189499},
journal = {IEEE Trans. Nucl. Sci.},
number = {6},
pages = {2822--2829},
title = {{Radiation-induced charge collection in infrared detector arrays}},
volume = {49 I},
year = {2002}
}

@article{Rauscher2014,
author = {Rauscher, Bernard J. and Boehm, Nicholas and Cagiano, Steve and Delo, Gregory S. and Foltz, Roger and Greenhouse, Matthew A. and Hickey, Michael and Hill, Robert J. and Kan, Emily and Lindler, Don and Mott, D. Brent and Waczynski, Augustyn and Wen, Yiting},
doi = {10.1086/677681},
issn = {00046280},
journal = {PASP},
number = {2008},
pages = {000--000},
title = {{New and Better Detectors for the JWST Near-Infrared Spectrograph}},
year = {2014}
}

@article{Rauscher_2017,
author = {Rauscher, Bernard J and Arendt, Richard G and Fixsen, D J and Greenhouse, Matthew A and Lander, Matthew and Lindler, Don and Loose, Markus and Moseley, S H and Mott, D Brent and Wen, Yiting and Wilson, Donna V and Xenophontos, Christos},
doi = {10.1088/1538-3873/aa83fd},
journal = {PASP},
month = {sep},
number = {980},
pages = {105003},
publisher = {The Astronomical Society of the Pacific},
title = {{Improved Reference Sampling and Subtraction: A Technique for Reducing the Read Noise of Near-infrared Detector Systems}},
url = {https://dx.doi.org/10.1088/1538-3873/aa83fd},
volume = {129},
year = {2017}
}

@article{Stone1998SSRv...86....1S,
    author         = {{Stone}, E.~C. and {Frandsen}, A.~M. and {Mewaldt}, R.~A. and {Christian}, E.~R. and {Margolies}, D. and {Ormes}, J.~F. and {Snow}, F.},
    title          = "{The Advanced Composition Explorer}",
    journal        = {Space Science Reviews},
    year           = {1998},
    month          = {Jun},
    volume         = {86},
    number         = {1-4},
    pages          = {1-22},
    doi            = {10.1023/A:1005082526237},
    adsurl         = {https://ui.adsabs.harvard.edu/abs/1998SSRv...86....1S}
}

@article{Gardner2023,
archivePrefix = {arXiv},
arxivId = {astro-ph.IM/2304.04869},
author = {Gardner, Jonathan P and Mather, John C and Abbott, Randy and Abell, James S and Abernathy, Mark and Abney, Faith E and Abraham, John G and Abraham, Roberto and Abul-Huda, Yasin M and Acton, Scott and Adams, Cynthia K and Adams, Evan and Adler, David S and Adriaensen, Maarten and Aguilar, Jonathan Albert and Ahmed, Mansoor and Ahmed, Nasif S and Ahmed, Tanjira and Albat, R{\"{u}}deger and Albert, Lo$\backslash$"$\backslash$ic and Alberts, Stacey and Aldridge, David and Allen, Mary Marsha and Allen, Shaune S and Altenburg, Martin and Altunc, Serhat and Alvarez, Jose Lorenzo and {\'{A}}lvarez-M{\'{a}}rquez, Javier and {Alves de Oliveira}, Catarina and Ambrose, Leslie L and Anandakrishnan, Satya M and Andersen, Gregory C and Anderson, Harry James and Anderson, Jay and Anderson, Kristen and Anderson, Sara M and Aprea, Julio and Archer, Benita J and Arenberg, Jonathan W and Argyriou, Ioannis and Arribas, Santiago and Artigau, {\'{E}}tienne and Arvai, Amanda Rose and Atcheson, Paul and Atkinson, Charles B and Averbukh, Jesse and Aymergen, Cagatay and Bacinski, John J and Baggett, Wayne E and Bagnasco, Giorgio and Baker, Lynn L and Balzano, Vicki Ann and Banks, Kimberly A and Baran, David A and Barker, Elizabeth A and Barrett, Larry K and Barringer, Bruce O and Barto, Allison and Bast, William and Baudoz, Pierre and Baum, Stefi and Beatty, Thomas G and Beaulieu, Mathilde and Bechtold, Kathryn and Beck, Tracy and Beddard, Megan M and Beichman, Charles and Bellagama, Larry and Bely, Pierre and Berger, Timothy W and Bergeron, Louis E and Bernier, Antoine-Darveau and Bertch, Maria D and Beskow, Charlotte and Betz, Laura E and Biagetti, Carl P and Birkmann, Stephan and Bjorklund, Kurt F and Blackwood, James D and Blazek, Ronald Paul and Blossfeld, Stephen and Bluth, Marcel and Boccaletti, Anthony and {Boegner Martin E.}, Jr. and Bohlin, Ralph C and Boia, John Joseph and B{\"{o}}ker, Torsten and Bonaventura, N and Bond, Nicholas A and Bosley, Kari Ann and Boucarut, Rene A and Bouchet, Patrice and Bouwman, Jeroen and Bower, Gary and Bowers, Ariel S and Bowers, Charles W and Boyce, Leslye A and Boyer, Christine T and Boyer, Martha L and Boyer, Michael and Boyer, Robert and Bradley, Larry D and Brady, Gregory R and Brandl, Bernhard R and Brannen, Judith L and Breda, David and Bremmer, Harold G and Brennan, David and Bresnahan, Pamela A and Bright, Stacey N and Broiles, Brian J and Bromenschenkel, Asa and Brooks, Brian H and Brooks, Keira J and Brown, Bob and Brown, Bruce and Brown, Thomas M and Bruce, Barry W and Bryson, Jonathan G and Bujanda, Edwin D and Bullock, Blake M and Bunker, A.{\~{}}J. and Bureo, Rafael and Burt, Irving J and Bush, James Aaron and Bushouse, Howard A and Bussman, Marie C and Cabaud, Olivier and Cale, Steven and Calhoon, Charles D and Calvani, Humberto and Canipe, Alicia M and Caputo, Francis M and Cara, Mihai and Carey, Larkin and Case, Michael Eli and Cesari, Thaddeus and Cetorelli, Lee D and Chance, Don R and Chandler, Lynn and Chaney, Dave and Chapman, George N and Charlot, S and Chayer, Pierre and Cheezum, Jeffrey I and Chen, Bin and Chen, Christine H and Cherinka, Brian and Chichester, Sarah C and Chilton, Zachary S and Chittiraibalan, Dharini and Clampin, Mark and Clark, Charles R and Clark, Kerry W and Clark, Stephanie M and Claybrooks, Edward E and Cleveland, Keith A and Cohen, Andrew L and Cohen, Lester M and Col{\'{o}}n, Knicole D and Coleman, Benee L and Colina, Luis and Comber, Brian J and Comeau, Thomas M and Comer, Thomas and {Conde Reis}, Alain and Connolly, Dennis C and Conroy, Kyle E and Contos, Adam R and Contreras, James and Cook, Neil J and Cooper, James L and Cooper, Rachel Aviva and Correia, Michael F and Correnti, Matteo and Cossou, Christophe and Costanza, Brian F and Coulais, Alain and Cox, Colin R and Coyle, Ray T and Cracraft, Misty M and Crew, Keith A and Curtis, Gary J and Cusveller, Bianca and {Da Costa Maciel}, Cleyciane and Dailey, Christopher T and Daugeron, Fr{\'{e}}d{\'{e}}ric and Davidson, Greg S and Davies, James E and Davis, Katherine Anne and Davis, Michael S and Day, Ratna and de Chambure, Daniel and de Jong, Pauline and {De Marchi}, Guido and Dean, Bruce H and Decker, John E and Delisa, Amy S and Dell, Lawrence C and Dellagatta, Gail and Dembinska, Franciszka and Demosthenes, Sandor and Dencheva, Nadezhda M and Deneu, Philippe and DePriest, William W and Deschenes, Jeremy and Dethienne, Nathalie and Detre, {\"{O}}rs Hunor and Diaz, Rosa Izela and Dicken, Daniel and DiFelice, Audrey S and Dillman, Matthew and Disharoon, Maureen O and Dixon, William V and Doggett, Jesse B and Dominguez, Keisha L and Donaldson, Thomas S and Doria-Warner, Cristina M and Santos, Tony Dos and Doty, Heather and {Douglas Robert E.}, Jr. and Doyon, Ren{\'{e}} and Dressler, Alan and Driggers, Jennifer and Driggers, Phillip A and Dunn, Jamie L and DuPrie, Kimberly C and Dupuis, Jean and Durning, John and Dutta, Sanghamitra B and Earl, Nicholas M and Eccleston, Paul and Ecobichon, Pascal and Egami, Eiichi and Ehrenwinkler, Ralf and Eisenhamer, Jonathan D and Eisenhower, Michael and Eisenstein, Daniel J and {El Hamel}, Zaky and Elie, Michelle L and Elliott, James and Elliott, Kyle Wesley and Engesser, Michael and Espinoza, N{\'{e}}stor and Etienne, Odessa and Etxaluze, Mireya and Evans, Leah and Fabreguettes, Luce and Falcolini, Massimo and Falini, Patrick R and Fatig, Curtis and Feeney, Matthew and Feinberg, Lee D and Fels, Raymond and Ferdous, Nazma and Ferguson, Henry C and Ferrarese, Laura and Ferreira, Marie-H{\'{e}}l{\'{e}}ne and Ferruit, Pierre and Ferry, Malcolm and Filippazzo, Joseph Charles and Firre, Daniel and Fix, Mees and Flagey, Nicolas and Flanagan, Kathryn A and Fleming, Scott W and Florian, Michael and Flynn, James R and Foiadelli, Luca and Fontaine, Mark R and Fontanella, Erin Marie and Forshay, Peter Randolph and Fortner, Elizabeth A and Fox, Ori D and Framarini, Alexandro P and Francisco, John I and Franck, Randy and Franx, Marijn and Franz, David E and Friedman, Scott D and Friend, Katheryn E and Frost, James R and Fu, Henry and Fullerton, Alexander W and Gaillard, Lionel and Galkin, Sergey and Gallagher, Ben and Galyer, Anthony D and {Garc$\backslash$'$\backslash$ia Mar$\backslash$'$\backslash$in}, Macarena and Gardner, Lisa E and Garland, Dennis and Garrett, Bruce Albert and Gasman, Danny and G{\'{a}}sp{\'{a}}r, Andr{\'{a}}s and Gastaud, Ren{\'{e}} and Gaudreau, Daniel and Gauthier, Peter Timothy and Geers, Vincent and Geithner, Paul H and Gennaro, Mario and Gerber, John and Gereau, John C and Giampaoli, Robert and Giardino, Giovanna and Gibbons, Paul C and Gilbert, Karoline and Gilman, Larry and Girard, Julien H and Giuliano, Mark E and Gkountis, Konstantinos and Glasse, Alistair and Glassmire, Kirk Zachary and Glauser, Adrian Michael and Glazer, Stuart D and Goldberg, Joshua and Golimowski, David A and Gonzaga, Shireen P and Gordon, Karl D and Gordon, Shawn J and Goudfrooij, Paul and Gough, Michael J and Graham, Adrian J and Grau, Christopher M and Green, Joel David and Greene, Gretchen R and Greene, Thomas P and Greenfield, Perry E and Greenhouse, Matthew A and Greve, Thomas R and Greville, Edgar M and Grimaldi, Stefano and Groe, Frank E and Groebner, Andrew and Grumm, David M and Grundy, Timothy and G{\"{u}}del, Manuel and Guillard, Pierre and Guldalian, John and Gunn, Christopher A and Gurule, Anthony and Gutman, Irvin Meyer and Guy, Paul D and Guyot, Benjamin and Hack, Warren J and Haderlein, Peter and Hagan, James B and Hagedorn, Andria and Hainline, Kevin and Haley, Craig and Hami, Maryam and Hamilton, Forrest Clifford and Hammann, Jeffrey and Hammel, Heidi B and Hanley, Christopher J and Hansen, Carl August and Hardy, Bruce and Harnisch, Bernd and Harr, Michael Hunter and Harris, Pamela and Hart, Jessica Ann and Hartig, George F and Hasan, Hashima and Hashim, Kathleen Marie and Hashimoto, Ryan and Haskins, Sujee J and Hawkins, Robert Edward and Hayden, Brian and Hayden, William L and Healy, Mike and Hecht, Karen and Heeg, Vince J and Hejal, Reem and Helm, Kristopher A and Hengemihle, Nicholas J and Henning, Thomas and Henry, Alaina and Henry, Ronald L and Henshaw, Katherine and Hernandez, Scarlin and Herrington, Donald C and Heske, Astrid and Hesman, Brigette Emily and Hickey, David L and Hilbert, Bryan N and Hines, Dean C and Hinz, Michael R and Hirsch, Michael and Hitcho, Robert S and Hodapp, Klaus and Hodge, Philip E and Hoffman, Melissa and Holfeltz, Sherie T and Holler, Bryan Jason and Hoppa, Jennifer Rose and Horner, Scott and Howard, Joseph M and Howard, Richard J and Huber, Jean M and Hunkeler, Joseph S and Hunter, Alexander and Hunter, David Gavin and Hurd, Spencer W and Hurst, Brendan J and Hutchings, John B and Hylan, Jason E and Ignat, Luminita Ilinca and Illingworth, Garth and Irish, Sandra M and {Isaacs John C.}, I I I and {Jackson Wallace C.}, Jr. and Jaffe, Daniel T and Jahic, Jasmin and Jahromi, Amir and Jakobsen, Peter and James, Bryan and James, John C and James, LeAndrea Rae and Jamieson, William Brian and Jandra, Raymond D and Jayawardhana, Ray and Jedrzejewski, Robert and Jeffers, Basil S and Jensen, Peter and Joanne, Egges and Johns, Alan T and Johnson, Carl A and Johnson, Eric L and Johnson, Patricia and Johnson, Phillip Stephen and Johnson, Thomas K and Johnson, Timothy W and Johnstone, Doug and Jollet, Delphine and Jones, Danny P and Jones, Gregory S and Jones, Olivia C and Jones, Ronald A and Jones, Vicki and Jordan, Ian J and Jordan, Margaret E and Jue, Reginald and Jurkowski, Mark H and Justis, Grant and Justtanont, Kay and Kaleida, Catherine C and Kalirai, Jason S and Kalmanson, Phillip Cabrales and Kaltenegger, Lisa and Kammerer, Jens and Kan, Samuel K and Kanarek, Graham Childs and Kao, Shaw-Hong and Karakla, Diane M and Karl, Hermann and Kassin, Susan A and Kauffman, David D and Kavanagh, Patrick and Kelley, Leigh L and Kelly, Douglas M and Kendrew, Sarah and Kennedy, Herbert V and Kenny, Deborah A and Keski-Kuha, Ritva A and Keyes, Charles D and Khan, Ali and Kidwell, Richard C and Kimble, Randy A and King, James S and King, Richard C and Kinzel, Wayne M and Kirk, Jeffrey R and Kirkpatrick, Marc E and Klaassen, Pamela and Klingemann, Lana and Klintworth, Paul U and Knapp, Bryan Adam and Knight, Scott and Knollenberg, Perry J and Knutsen, Daniel Mark and Koehler, Robert and Koekemoer, Anton M and Kofler, Earl T and Kontson, Vicki L and Kovacs, Aiden Rose and Kozhurina-Platais, Vera and Krause, Oliver and Kriss, Gerard A and Krist, John and Kristoffersen, Monica R and Krogel, Claudia and Krueger, Anthony P and Kulp, Bernard A and Kumari, Nimisha and Kwan, Sandy W and Kyprianou, Mark and Labador, Aurora Gadiano and Labiano, {\'{A}}lvaro and Lafreni{\`{e}}re, David and Lagage, Pierre-Olivier and Laidler, Victoria G and Laine, Benoit and Laird, Simon and Lajoie, Charles-Philippe and Lallo, Matthew D and Lam, May Yen and LaMassa, Stephanie Marie and Lambros, Scott D and Lampenfield, Richard Joseph and Lander, Matthew Ed and Langston, James Hutton and Larson, Kirsten and Larson, Melora and LaVerghetta, Robert Joseph and Law, David R and Lawrence, Jon F and Lee, David W and Lee, Janice and Lee, Yat-Ning Paul and Leisenring, Jarron and Leveille, Michael Dunlap and Levenson, Nancy A and Levi, Joshua S and Levine, Marie B and Lewis, Dan and Lewis, Jake and Lewis, Nikole and Libralato, Mattia and Lidon, Norbert and Liebrecht, Paula Louisa and Lightsey, Paul and Lilly, Simon and Lim, Frederick C and Lim, Pey Lian and Ling, Sai-Kwong and Link, Lisa J and Link, Miranda Nicole and Lipinski, Jamie L and Liu, XiaoLi and Lo, Amy S and Lobmeyer, Lynette and Logue, Ryan M and Long, Chris A and Long, Douglas R and Long, Ilana D and Long, Knox S and L{\'{o}}pez-Caniego, Marcos and Lotz, Jennifer M and Love-Pruitt, Jennifer M and Lubskiy, Michael and Luers, Edward B and Luetgens, Robert A and Luevano, Annetta J and Lui, Sarah Marie G Flores and {Lund James M.}, I I I and Lundquist, Ray A and Lunine, Jonathan and L{\"{u}}tzgendorf, Nora and Lynch, Richard J and MacDonald, Alex J and MacDonald, Kenneth and Macias, Matthew J and Macklis, Keith I and Maghami, Peiman and Maharaja, Rishabh Y and Maiolino, Roberto and Makrygiannis, Konstantinos G and Malla, Sunita Giri and Malumuth, Eliot M and Manjavacas, Elena and Marini, Andrea and Marrione, Amanda and Marston, Anthony and Martel, Andr{\'{e}} R and Martin, Didier and Martin, Peter G and Martinez, Kristin L and Maschmann, Marc and Masci, Gregory L and Masetti, Margaret E and Maszkiewicz, Michael and Matthews, Gary and Matuskey, Jacob E and McBrayer, Glen A and McCarthy, Donald W and McCaughrean, Mark J and McClare, Leslie A and McClare, Michael D and McCloskey, John C and McClurg, Taylore D and McCoy, Martin and McElwain, Michael W and McGregor, Roy D and McGuffey, Douglas B and McKay, Andrew G and McKenzie, William K and McLean, Brian and McMaster, Matthew and McNeil, Warren and {De Meester}, Wim and Mehalick, Kimberly L and Meixner, Margaret and Mel{\'{e}}ndez, Marcio and Menzel, Michael P and Menzel, Michael T and Merz, Matthew and Mesterharm, David D and Meyer, Michael R and Meyett, Michele L and Meza, Luis E and Midwinter, Calvin and Milam, Stefanie N and Miller, Jay Todd and Miller, William C and Miskey, Cherie L and Misselt, Karl and Mitchell, Eileen P and Mohan, Martin and Montoya, Emily E and Moran, Michael J and Morishita, Takahiro and Moro-Mart$\backslash$'$\backslash$in, Amaya and Morrison, Debra L and Morrison, Jane and Morse, Ernie C and Moschos, Michael and Moseley, S.{\~{}}H. and Mosier, Gary E and Mosner, Peter and Mountain, Matt and Muckenthaler, Jason S and Mueller, Donald G and Mueller, Migo and Muhiem, Daniella and M{\"{u}}hlmann, Prisca and Mullally, Susan Elizabeth and Mullen, Stephanie M and Munger, Alan J and Murphy, Jess and Murray, Katherine T and Muzerolle, James C and Mycroft, Matthew and Myers, Andrew and Myers, Carey R and Myers, Fred Richard R and Myers, Richard and Myrick, Kaila and {Nagle Adrian F.}, I V and Nayak, Omnarayani and Naylor, Bret and Neff, Susan G and Nelan, Edmund P and Nella, John and Nguyen, Duy Tuong and Nguyen, Michael N and Nickson, Bryony and Nidhiry, John Joseph and Niedner, Malcolm B and Nieto-Santisteban, Maria and Nikolov, Nikolay K and Nishisaka, Mary Ann and Noriega-Crespo, Alberto and Nota, Antonella and O'Mara, Robyn C and Oboryshko, Michael and O'Brien, Marcus B and Ochs, William R and Offenberg, Joel D and Ogle, Patrick Michael and Ohl, Raymond G and Olmsted, Joseph Hamden and Osborne, Shannon Barbara and O'Shaughnessy, Brian Patrick and {\"{O}}stlin, G{\"{o}}ran and O'Sullivan, Brian and Otor, O Justin and Ottens, Richard and Ouellette, Nathalie N -Q. and Outlaw, Daria J and Owens, Beverly A and Pacifici, Camilla and Page, James Christophe and Paranilam, James G and Park, Sang and Parrish, Keith A and Paschal, Laura and Patapis, Polychronis and Patel, Jignasha and Patrick, Keith and {Pattishall Robert A.}, Jr. and Paul, Douglas William and Paul, Shirley J and Pauly, Tyler Andrew and Pavlovsky, Cheryl M and Pe{\~{n}}a-Guerrero, Maria and Pedder, Andrew H and Peek, Matthew Weldon and Pelham, Patricia A and Penanen, Konstantin and Perriello, Beth A and Perrin, Marshall D and Perrine, Richard F and Perrygo, Chuck and Peslier, Muriel and Petach, Michael and Peterson, Karla A and Pfarr, Tom and Pierson, James M and Pietraszkiewicz, Martin and Pilchen, Guy and Pipher, Judy L and Pirzkal, Norbert and Pitman, Joseph T and Player, Danielle M and Plesha, Rachel and Plitzke, Anja and Pohner, John A and Poletis, Karyn Konstantin and Pollizzi, Joseph A and Polster, Ethan and Pontius, James T and Pontoppidan, Klaus and Porges, Susana C and Potter, Gregg D and Prescott, Stephen and Proffitt, Charles R and Pueyo, Laurent and {Quispe Neira}, Irma Aracely and Radich, Armando and Rager, Reiko T and Rameau, Julien and Ramey, Deborah D and {Ramos Alarcon}, Rafael and Rampini, Riccardo and Rapp, Robert and Rashford, Robert A and Rauscher, Bernard J and Ravindranath, Swara and Rawle, Timothy and Rawlings, Tynika N and Ray, Tom and Regan, Michael W and Rehm, Brian and Rehm, Kenneth D and Reid, Neill and Reis, Carl A and Renk, Florian and Reoch, Tom B and Ressler, Michael and Rest, Armin W and Reynolds, Paul J and Richon, Joel G and Richon, Karen V and Ridgaway, Michael and Riedel, Adric Richard and Rieke, George H and Rieke, Marcia J and Rifelli, Richard E and Rigby, Jane R and Riggs, Catherine S and Ringel, Nancy J and Ritchie, Christine E and Rix, Hans-Walter and Robberto, Massimo and Robinson, Gregory L and Robinson, Michael S and Robinson, Orion and Rock, Frank W and Rodriguez, David R and {Rodr$\backslash$'$\backslash$iguez del Pino}, Bruno and Roellig, Thomas and Rohrbach, Scott O and Roman, Anthony J and Romelfanger, Frederick J and {Romo Felipe P.}, Jr. and Rosales, Jose J and Rose, Perry and Roteliuk, Anthony F and Roth, Marc N and Rothwell, Braden Quinn and Rouzaud, Sylvain and Rowe, Jason and Rowlands, Neil and Roy, Arpita and Royer, Pierre and Rui, Chunlei and Rumler, Peter and Rumpl, William and Russ, Melissa L and Ryan, Michael B and Ryan, Richard M and Saad, Karl and Sabata, Modhumita and Sabatino, Rick and Sabbi, Elena and Sabelhaus, Phillip A and Sabia, Stephen and Sahu, Kailash C and Saif, Babak N and Salvignol, Jean-Christophe and Samara-Ratna, Piyal and Samuelson, Bridget S and Sanders, Felicia A and Sappington, Bradley and Sargent, B.{\~{}}A. and Sauer, Arne and Savadkin, Bruce J and Sawicki, Marcin and Schappell, Tina M and Scheffer, Caroline and Scheithauer, Silvia and Scherer, Ron and Schiff, Conrad and Schlawin, Everett and Schmeitzky, Olivier and Schmitz, Tyler S and Schmude, Donald J and Schneider, Analyn and Schreiber, J{\"{u}}rgen and Schroeven-Deceuninck, Hilde and Schultz, John J and Schwab, Ryan and Schwartz, Curtis H and Scoccimarro, Dario and Scott, John F and Scott, Michelle B and Seaton, Bonita L and Seely, Bruce S and Seery, Bernard and Seidleck, Mark and Sembach, Kenneth and Shanahan, Clare Elizabeth and Shaughnessy, Bryan and Shaw, Richard A and Shay, Christopher Michael and Sheehan, Even and Sheth, Kartik and Shih, Hsin-Yi and Shivaei, Irene and Siegel, Noah and Sienkiewicz, Matthew G and Simmons, Debra D and Simon, Bernard P and Sirianni, Marco and Sivaramakrishnan, Anand and Slade, Jeffrey E and Sloan, G.{\~{}}C. and Slocum, Christine E and Slowinski, Steven E and Smith, Corbett T and Smith, Eric P and Smith, Erin C and Smith, Koby and Smith, Robert and Smith, Stephanie J and Smolik, John L and Soderblom, David R and Sohn, Sangmo Tony and Sokol, Jeff and Sonneborn, George and Sontag, Christopher D and Sooy, Peter R and Soummer, Remi and Southwood, Dana M and Spain, Kay and Sparmo, Joseph and Speer, David T and Spencer, Richard and Sprofera, Joseph D and Stallcup, Scott S and Stanley, Marcia K and Stansberry, John A and Stark, Christopher C and Starr, Carl W and Stassi, Diane Y and Steck, Jane A and Steeley, Christine D and Stephens, Matthew A and Stephenson, Ralph J and Stewart, Alphonso C and Stiavelli, Massimo and Stockman, Hervey Jr. and Strada, Paolo and Straughn, Amber N and Streetman, Scott and Strickland, David Kendal and Strobele, Jingping F and Stuhlinger, Martin and Stys, Jeffrey Edward and Such, Miguel and Sukhatme, Kalyani and Sullivan, Joseph F and Sullivan, Pamela C and Sumner, Sandra M and Sun, Fengwu and Sunnquist, Benjamin Dale and Swade, Daryl Allen and Swam, Michael S and Swenton, Diane F and Swoish, Robby A and {Tam Litten}, Oi In and Tamas, Laszlo and Tao, Andrew and Taylor, David K and Taylor, Joanna M and te Plate, Maurice and {Van Tea}, Mason and Teague, Kelly K and Telfer, Randal C and Temim, Tea and Texter, Scott C and Thatte, Deepashri G and Thompson, Christopher Lee and Thompson, Linda M and Thomson, Shaun R and Thronson, Harley and Tierney, C.{\~{}}M. and Tikkanen, Tuomo and Tinnin, Lee and Tippet, William Thomas and Todd, Connor William and Tran, Hien D and Trauger, John and Trejo, Edwin Gregorio and {Vinh Truong}, Justin Hoang and Tsukamoto, Christine L and Tufail, Yasir and Tumlinson, Jason and Tustain, Samuel and Tyra, Harrison and Ubeda, Leonardo and Underwood, Kelli and Uzzo, Michael A and Vaclavik, Steven and Valenduc, Frida and Valenti, Jeff A and {Van Campen}, Julie and van de Wetering, Inge and {Van Der Marel}, Roeland P and van Haarlem, Remy and Vandenbussche, Bart and van Dishoeck, Ewine F and Vanterpool, Dona D and Vernoy, Michael R and {Vila Costas}, Maria Bego{\~{n}}a and Volk, Kevin and Voorzaat, Piet and Voyton, Mark F and Vydra, Ekaterina and Waddy, Darryl J and Waelkens, Christoffel and Wahlgren, Glenn Michael and {Walker Frederick E.}, Jr. and Wander, Michel and Warfield, Christine K and Warner, Gerald and Wasiak, Francis C and Wasiak, Matthew F and Wehner, James and Weiler, Kevin R and Weilert, Mark and Weiss, Stanley B and Wells, Martyn and Welty, Alan D and Wheate, Lauren and Wheeler, Thomas P and White, Christy L and Whitehouse, Paul and Whiteleather, Jennifer Margaret and Whitman, William Russell and Williams, Christina C and Willmer, Christopher N.{\~{}}A. and Willott, Chris J and Willoughby, Scott P and Wilson, Andrew and Wilson, Debra and Wilson, Donna V and Windhorst, Rogier and Wislowski, Emily Christine and Wolfe, David J and Wolfe, Michael A and Wolff, Schuyler and Wondel, Amancio and Woo, Cindy and Woods, Robert T and Worden, Elaine and Workman, William and Wright, Gillian S and Wu, Carl and Wu, Chi-Rai and Wun, Dakin D and Wymer, Kristen B and Yadetie, Thomas and Yan, Isabelle C and Yang, Keith C and Yates, Kayla L and Yeager, Christopher R and Yerger, Ethan John and Young, Erick T and Young, Gary and Yu, Gene and Yu, Susan and Zak, Dean S and Zeidler, Peter and Zepp, Robert and Zhou, Julia and Zincke, Christian A and Zonak, Stephanie and Zondag, Elisabeth},
doi = {10.1088/1538-3873/acd1b5},
eprint = {2304.04869},
journal = {PASP},
month = {jun},
number = {1048},
pages = {68001},
primaryClass = {astro-ph.IM},
title = {{The James Webb Space Telescope Mission}},
volume = {135},
year = {2023}
}
\bibliographystyle{aasjournal}

\end{document}